\documentclass[aps,prd,superscriptaddress,twocolumn,nofootinbib,10pt]{revtex4}
\usepackage[pdftex]{color}
\usepackage[sort&compress]{natbib}
\usepackage[colorlinks=true,linkcolor=blue,filecolor=blue,urlcolor=blue,citecolor=blue,pdftex,plainpages=false]{hyperref}
\usepackage{url}
\pdfoutput=1
\usepackage{ifpdf}
\usepackage{color,hhline,psfrag,rotating}
\usepackage{dcolumn}
\usepackage{verbatim}
\usepackage{bm}
\usepackage{graphicx}
\usepackage{amsfonts,amssymb,amsmath}
\usepackage{booktabs}  
\usepackage{array}

\newcommand{\lqcd}{\Lambda_{\text{QCD}}}

\newcommand{\Xsl}[1]{\raise.15ex\hbox{/}\kern-.57em #1}
\newcommand{\order}[1]{\mathcal{O}(#1)}

\begin{document}
\title{Lattice QCD form factor for $B_s\to D_s^* \ell\nu$ at zero recoil with non-perturbative current renormalisation}

\author{E.~McLean}
\email[]{e.mclean.1@research.gla.ac.uk}
\affiliation{SUPA, School of Physics and Astronomy, University of Glasgow, Glasgow, G12 8QQ, UK}
\author{C.~T.~H.~Davies}
\email[]{christine.davies@glasgow.ac.uk}
\affiliation{SUPA, School of Physics and Astronomy, University of Glasgow, Glasgow, G12 8QQ, UK}
\author{A.~T.~Lytle}
\affiliation{INFN, Sezione di Roma Tor Vergata, Via della Ricerca Scientifica 1, 00133 Roma RM, Italy}
\author{J.~Koponen}
\affiliation{High Energy Accelerator Research Organisation (KEK), Tsukuba 305-0801, Japan}
\collaboration{HPQCD collaboration}
\homepage{http://www.physics.gla.ac.uk/HPQCD}
\noaffiliation

\date{\today}

\begin{abstract}
  We present details of a lattice QCD calculation of the $B_s\to D_s^*$ axial form factor at zero recoil using 
the Highly Improved Staggered Quark (HISQ) formalism on the second generation MILC gluon ensembles that include 
up, down, strange and charm quarks in the sea. 
Using the HISQ action for all valence quarks means that the lattice 
axial vector current that couples to the $W$ can be renormalized fully non-perturbatively, 
giving a result free of the perturbative matching errors that previous lattice QCD calculations have 
had. We calculate correlation functions at three values of the lattice spacing, and multiple
`$b$'-quark masses, for physical $c$ and $s$.
The functional dependence on the $b$-quark mass can be determined and compared to Heavy Quark 
Effective Theory expectations, and a result for the form factor obtained at 
the physical value of the $b$-quark mass. 
We find $\mathcal{F}^{B_s\to D_s^*}(1) = h^s_{A_1}(1) = 0.9020(96)_{\text{stat}}(90)_{\text{sys}}$. 
This is in agreement with earlier lattice QCD results, which use NRQCD $b$ quarks, 
with a total uncertainty reduced 
by more than a factor of two. We discuss implications of this result 
for the $B\to D^*$ axial form factor at zero recoil and for determinations
of $V_{cb}$.
\end{abstract}

\maketitle
\section{Introduction}
\label{sec:intro}

The study of quark flavour-changing interactions is a key component of the search for 
physics beyond the Standard Model (SM). There are currently a number of related tensions 
between experimental measurements and SM predictions~\cite{Wei:2009zv,Lees:2012xj,Lees:2012tva,Lees:2013uzd,Aaij:2014pli,Aaij:2014ora,Huschle:2015rga,Aaij:2015oid,Aaij:2015yra,Aaij:2015xza,Aaij:2016flj,Wehle:2016yoi,Sato:2016svk,Hirose:2017dxl,Aaij:2017deq,Aaij:2017uff,Aaij:2017vbb,Sirunyan:2017dhj,Aaboud:2018krd},
along with discrepancies between systematically independent determinations of 
Cabibbo-Kobayashi-Maskawa (CKM) matrix elements~\cite{Amhis:2016xyh,Bevan:2014iga,Alberti:2014yda}. 
A more precise understanding of these processes is needed to resolve these issues.

The $\bar{B}^0\to D^{*+} \ell^- \bar{\nu}$ decay (and its charge conjugate, 
that we simply abbreviate to $B \to D^* \ell \nu$ from now on) 
supplies one of the three methods used 
for precisely determining the CKM element $|V_{cb}|$~\cite{Schroder:1994aj,Bortoletto:1990fx,Fulton:1990bx,Albrecht:1991iz,Barish:1994mu,Buskulic:1996sm,Buskulic:1994dz,Abbiendi:2000hk,Abreu:2001ic,Adam:2002uw,Abdallah:2004rz,Aubert:2007rs,Aubert:2007qs,Aubert:2008yv,Dungel:2010uk,Abdesselam:2017kjf,Bailey:2014tva,Abdesselam:2018nnh}. Measurements of branching fractions are extrapolated through $q^2$ to the zero recoil point to deduce $\mathcal{F}(1) |V_{cb}|$, where $\mathcal{F}(1)$ is the value of the 
only form factor contributing at zero recoil. Then a determination of $\mathcal{F}(1)$ in the Standard Model (via Lattice QCD \cite{Bailey:2014tva,Harrison:2017fmw}) can be divided out to infer $|V_{cb}|$.

$|V_{cb}|$ is an important quantity and needs to be determined accurately. 
It constrains one side of the unitarity triangle via 
the ratio $|V_{ub}|/|V_{cb}|$. It is also a dominant uncertainty in the determination of 
the $CP$-violation parameter $\epsilon_K$ (where there is currently tension between 
the SM and experiment, see for example \cite{Bailey:2018feb}).  

Previous determinations of $|V_{cb}|$ have shown systematic discrepancies with each other. 
The two competing values were those derived from {\it{exclusive}} decays 
($B\to D^* \ell \nu$ and $B\to D \ell \nu$ with 
$B \rightarrow D^*$ giving the more accurare result), 
and {\it{inclusive}} ($B \to X_c \ell \nu$, where $X_c$ is any charmed 
hadronic state). In 2016 the Heavy Flavour Averaging Group (HFLAV) gave a 
value derived from 
exclusive $B \rightarrow D^*$ decays of $|V_{cb}|_{\text{excl}}= (39.05\pm 0.47_{\text{exp}}\pm 0.58_{\text{th}})\times10^{-3}$ and from 
inclusive decays, using the kinetic scheme, of 
$|V_{cb}|_{\text{incl}} = (42.19\pm 0.78)\times 10^{-3}$~\cite{Amhis:2016xyh}. 
It has since been suggested, based on unfolded Belle data~\cite{Abdesselam:2017kjf}, 
that the tension seen here arose 
(at least partly) from the use of a very constrained parameterization in the extrapolation 
of the experimental $B \rightarrow D^*$ decay rates 
to zero recoil \cite{Bernlochner:2017jka, Bigi:2017njr, Grinstein:2017nlq}. 
Recent exclusive determinations of $V_{cb}$ have then used a less constrained parameterisation 
to give a larger, and less precise, result for $V_{cb}$ that is no longer in tension with 
the inclusive result. For example, the Particle Data Group quote 
$|V_{cb}|_{\text{excl}}= (41.9\pm 2.0)\times10^{-3}$~\cite{Tanabashi:2018oca}. 
However, an even more recent $V_{cb}$ determination 
from $B\to D^*\ell \nu$ data by the BaBar collaboration~\cite{Dey:2019bgc} 
used the less constrained parameterisation but still found a tension with the inclusive result.
This clearly points to the need for more work to improve the accuracy of the exclusive result.
On the theory side a better understanding of the form 
factors for $B \rightarrow D^*$ from lattice QCD is required, both at zero recoil and away 
from zero recoil. 

Another motivation for studying $B\to D^* \ell \nu$ is 
the tension between SM and experimental determinations of the 
ratio $R_{D^{(*)}} = \mathcal{B}(\bar{B}\to D^{(*)} \tau \bar{\nu}_{\tau}) / \mathcal{B}(\bar{B}\to D^{(*)} \ell \bar{\nu}_{\ell})$ ($\ell=e$ or $\mu$). 
The latest HFLAV report gives the combined statistical significance 
of the anomalies in $R_D$ and $R_{D^*}$ to be 
$3.8\sigma$~\cite{Amhis:2016xyh}. A preliminary new analysis 
from Belle~\cite{CariaMoriond}, however, gives results closer to the 
SM and pulls the global average down to $3.1\sigma$.
More precise measurements and predictions will either confirm or dismiss 
a new physics explanation.

The  weak decay process $B_s \to D^*_s \ell \nu$ is very 
similar to $B\to D^*\ell \nu$ and could also be used to determine $|V_{cb}|$ and test 
the SM. It is feasible to study this decay at LHC and from the theoretical side it is a 
more attractive channel than $B \to D^*$.  
The absence of valence light quarks means that lattice QCD results have 
smaller statistical errors and are less computationally expensive. Finite-volume effects 
and the dependence on $u/d$ quark masses (for quarks in the sea) are also smaller. 
The $D_s^*$ has no Zweig-allowed strong decay mode, unlike the $D^*$, and is in 
fact a relatively long-lived particle~\cite{Donald:2013sra} that can be considered 
`gold-plated' in lattice QCD.  
This makes the $B_s \to D^*_s \ell\nu$ both a useful test bed 
for lattice techniques (that may be later used to study 
$B \to D^* \ell \nu$ decays) and a key decay process for which 
to make predictions ahead of experimental results.

Lattice QCD calculations have shown that several weak decay form factors are relatively 
insensitive to whether the spectator quark is a $u/d$ or $s$ quark~\cite{Bailey:2012rr, Koponen:2013tua, Monahan:2017uby}.
A combination of chiral perturbation theory and Heavy Quark Symmetry~\cite{Jenkins:1992qv} backs up 
this expectation for $B$ decays. 
We can therefore expect the form factors to be very similar for $B_s \to D^*_s$ 
and $B \to D^*$. 
A recent lattice calculation~\cite{Harrison:2017fmw} found an insignificant $\mathcal{O}(1\%)$ difference at zero recoil: 
$\mathcal{F}^{B\to D^*}(1) / \mathcal{F}^{B_s\to D_s^*}(1) = 1.013(14)_{\text{stat}}(17)_{\text{sys}}$. 
Information from the study of $B_s \to D_s^*$ can then be 
applicable to $B \to D^*$. 

Lattice QCD calculations of the $B_{(s)} \to D_{(s)}^*$ form factors at zero recoil 
have so far been performed by two collaborations using different methods. 
The Fermilab Lattice and MILC collaborations calculated $\mathcal{F}^{B\to D^*}(1)$ 
in~\cite{Bernard:2008dn, Bailey:2014tva} using the Fermilab action for both 
$b$ and $c$ quarks~\cite{ElKhadra:1996mp} and asqtad $u/d$ quarks~\cite{Bazavov:2009bb}. 
More recently the HPQCD collaboration 
computed both $\mathcal{F}^{B\to D^*}(1)$ and 
$\mathcal{F}^{B_{s}\to D_{s}^*}(1)$~\cite{Harrison:2017fmw} using 
improved NRQCD $b$ quarks~\cite{Lepage:1992tx, Dowdall:2011wh} and Highly Improved 
Staggered (HISQ) $c$ and $u/d/s$ quarks~\cite{Follana:2006rc}. 
The RBC/UKQCD~\cite{Flynn:2016vej} and LANL-SWME~\cite{Bailey:2017xjk} collaborations 
are also working towards these form factors using variants of the Fermilab action for 
heavy quarks and JLQCD has a calculation in progress using M\"{o}bius domain-wall
quarks~\cite{Kaneko:2018mcr}.

The formalism to use for the heavy quarks is a major consideration 
in designing a lattice QCD calculation to determine these form factors. 
Most of the calculations discussed in the previous paragraph (apart 
from the JLQCD calculation) use approaches that 
make use of the nonrelativistic nature of heavy quark bound states to tune 
the $b$ (and in some cases also $c$) quark masses. This avoids potentially 
large discretisation effects appearing in the results in the form of a 
systematic error of size $(am_b)^n$, where $n$ is an integer that depends on 
the level of improvement in the action. The absence of these discretisation 
errors means that $b$ quarks can be handled on relatively coarse lattices 
where $am_b > 1$. However the price to be paid is that the current operator 
that couples to the $W$ boson is also implemented within a nonrelativistic framework 
and must then be renormalised to match the appropriate operator in continuum QCD. 
This matching can be done using perturbative lattice QCD but has only be 
done through $\mathcal{O}(\alpha_s)$ for these actions~\cite{Harada:2001fj, Monahan:2012dq}. 
This leaves a substantial source of uncertainty from missing higher-order terms 
in the perturbative matching that is not easily reduced. 
This matching uncertainty contributes $\sim$ 80\% of the final error in 
the HPQCD calculation~\cite{Harrison:2017fmw} and $\sim$ 30\% in the 
Fermilab/MILC calculation~\cite{Bailey:2014tva} because of the differing allowances 
for missing higher-order terms. 

Here we report details and results of a calculation of the $B_s\to D^*_s$ form 
factor at zero recoil using an approach free of perturbative matching uncertainties. 
We perform our calculation on the second-generation MILC 
ensembles~\cite{Bazavov:2010ru, Bazavov:2012xda}, including effects from 2+1+1 
flavours in the sea using the HISQ action. 
We also use the HISQ action for all valence quarks. We obtain results 
at a number of differing masses for the $b$ (we refer to this generically 
as the {\it{heavy quark}} $h$), and perform an extrapolation to $m_h=m_b$. 
By using only HISQ quarks, we can obtain the normalizations of all required currents 
fully non-perturbatively. We refer to this as the {\it{heavy-HISQ}} approach. 
By using many heavy masses and multiple values of the lattice spacing, including 
very fine lattices, we can model both the form factor dependence on the heavy 
mass, and the discretisation effects associated with using large $am_h$ values.

The heavy-HISQ approach was developed by HPQCD to compute $B$ meson masses 
and decay constants~\cite{McNeile:2011ng,McNeile:2012qf} and the $b$ quark 
mass~\cite{McNeile:2010ji, Chakraborty:2014aca}. It is also now being used 
by other collaborations for these calculations~\cite{Bazavov:2017lyh, Petreczky:2019ozv}.  
A proof-of-principle application of heavy-HISQ to form factors was given 
in~\cite{Lytle:2016ixw, Colquhoun:2016osw} for $B_c$ decays, 
showing that the full $q^2$ range of the 
decay could be covered. 
Here we extend the approach to form factors for $B_s$ decays 
but working only at zero recoil, a straightforward extension of earlier 
work. 
Using the heavy-HISQ approach also has the added benefit of 
eludicating the dependence 
of form factors on heavy quark masses, meaning that we can test expectations 
from Heavy Quark Effective Theory (HQET). 

This article is structured in the following way: Section \ref{sec:lattice} defines the form 
factor and gives details of the lattice calculation, including the nonperturbative normalisation 
and extrapolation in heavy-quark mass; 
Section \ref{sec:results} presents our results and compares to earlier calculations and Section \ref{sec:conclusions} gives our conclusions and outlook. In the appendix, we give details of a number of tests we performed on the correlator fits and the continuum, 
chiral and heavy-quark extrapolations.

\section{Calculation Details}
\label{sec:lattice}

\subsection{Form Factors}

The differential decay rate for the $\bar{B}_s^0\to D_s^{*+} l^- \bar{\nu}_l$ decay is given in the SM by
\begin{align}
  {d \Gamma \over dw} &(\bar{B}_s^0\to D_s^{*+} l^- \bar{\nu}_l) = {G_F^2 M_{D_s^*}^3 | \bar{\eta}_{\text{EW}} V_{cb} |^2 \over 4\pi^3}
\\  &\times(M_{B_s}^2-M_{D_s^*}^2) \sqrt{w^2-1} \chi(w) | \mathcal{F}^{B_s\to D_s^*}(w) |^2. \nonumber
\end{align}
where $w = v_{B_s} \cdot v_{D^*_s}$, $v = p/M$ is the 4-velocity 
of  each meson, 
and $\chi(w)$ is a known function of $w$ with $\chi(1)=1$ 
(see, for example, appendix G of \cite{Harrison:2017fmw}). 
$\bar{\eta}_{\text{EW}}$ accounts for electroweak corrections from diagrams where 
photons or $Z$s are exchanged in addition to a $W^-$, as well as the Coulomb 
attraction of the final-state charged 
particles~\cite{Sirlin:1981ie,Ginsberg:1968pz,Atwood:1989em}. 
The differential decay rate for the $B_s^0\to D_s^{*-} l^+ \bar{\nu}_l$ is identical.

The form factor $\mathcal{F}^{B_s\to D_s^*}(w)$ is a linear combination 
of hadronic form factors that parameterize the vector and axial-vector matrix 
elements between initial and final state hadrons. 
A common choice of parameterization used in the context of Heavy Quark Effective 
Theory (HQET) is~\cite{Richman:1995wm}
\begin{align}
  \langle D^*_s(\epsilon)| V^{\mu} | B_s \rangle &= i \sqrt{M_{B_s}M_{D^*_s}} h^s_V(w) \epsilon_{\mu\nu\alpha\beta} \,\epsilon^{*\nu} v_{D_s^*}^{\alpha} v_{B_s}^{\beta}, \\
  \langle D^*_s(\epsilon)| A^{\mu} | B_s \rangle &= \sqrt{M_{B_s}M_{D^*_s}} [ h^s_{A_1}(w) (w+1) \epsilon^*_{\mu} - \\ \nonumber
    h^s_{A_2}(w)& \,\epsilon^*\cdot v_{B_s} v_{B_s\,\mu} - h^s_{A_3}(w) \,\epsilon^*\cdot v_{B_s} v_{D^*_s\,\mu} ],
\end{align}
where $V^{\mu} = \bar{c} \gamma^{\mu} b$ is the vector $b\to c$ current 
and $A^{\mu} = \bar{c} \gamma^{\mu} \gamma^5 b$ is the axial-vector current. 
$\epsilon$ is the polarization 4-vector of the $D_s^*$ final state.

At zero recoil ($w=1$), the vector matrix element vanishes, the axial-vector element simplifies to
\begin{align}
  \langle D^*_s(\epsilon)| A^{\mu} | B_s \rangle &= 2 \sqrt{M_{B_s}M_{D^*_s}} h^s_{A_1}(1) \epsilon^{*\,\mu},
\end{align}
and $\mathcal{F}^{B_s\to D_s^*}(w)$ reduces to
\begin{align}
  \mathcal{F}^{B_s\to D_s^*}(1) = h^s_{A_1}(1).
\end{align}
Our goal is to compute $h^s_{A_1}(1)$.

All we need to do this is the matrix element $\langle D^*_s(\epsilon)| A^{\mu} | B_s \rangle$ 
with both the $B_s$ and $D_s^*$ at rest, with the $D_s^*$ polarization $\epsilon$ in the 
same direction as the (spatial) axial-vector current.

\subsection{Lattice Calculation}

The gluon field configurations that we use were generated by the MILC 
collaboration~\cite{Bazavov:2010ru,Bazavov:2012xda}. 
Table~\ref{tab:ensembles} gives the relevant parameters for the specific ensembles 
that we use. The gluon field is generated using a Symanzik-improved gluon action 
with coefficients calculated through $\order{\alpha_s a^2,n_f\alpha_s a^2}$~\cite{Hart:2008sq}. 
The configurations include the effect of 2+1+1 flavours of dynamical quarks 
in the sea ($u$,$d$,$s$,$c$, with $m_u=m_d\equiv m_l$), using the HISQ action~\cite{Follana:2006rc}. 
In three of the four ensembles (fine, superfine and ultrafine), the bare light quark 
mass is set to $m_{l0}/m_{s0} = 0.2$. 
The fact that the $m_{l0}$ value is unphysically high is expected to have only a 
small effect on $h^s_{A_1}(1)$, because there are no valence light quarks. 
The effect is quantified here by including a fourth ensemble (fine-physical) 
with (approximately) physical $m_{l0}$.

We use a number of different masses for the valence heavy quark, $h$. 
This is in order to resolve the dependence of $h_{A_1}^s(1)$ 
on the heavy mass, so that an 
extrapolation to $m_h=m_b$ can be performed. By varying the heavy mass on each ensemble 
and by using ensembles at varying small lattice spacing, 
we can resolve both the discretisation effects that grow with 
heavy quark mass ($am^{\text{val}}_{h0} \lesssim 1$) and the physical 
dependence of the continuum form factor on $m_h$.

\begin{table*}[t]
  \begin{center}
    \begin{tabular}{c c c c c c c c c c c c}
      \hline
      set & handle & $w_0/a$  & $N_x^3\times N_t$ & $n_{\text{cfg}}\times n_{\text{src}}$ & $am_{l0}$ & $am_{s0}$ & $am_{c0}$ & $am_{s0}^{\text{val}}$ & $am_{c0}^{\text{val}}$ & $am^{\text{val}}_{h0}$ & T \\ [0.5ex]
      \hline
      1 & \bf{fine} & 1.9006(20) & $32^3\times96$ & $938 \times 8$ & 0.0074 & 0.037 & 0.440 & 0.0376 & 0.45 
      & 0.5, 0.65, 0.8 & 14,17,20 \\ [1ex]
      2 & \bf{fine-physical} & 1.9518(7) & $64^3\times96$ & $284 \times 4$  & 0.0012 & 0.0363 & 0.432 & 0.036 & 0.433 
      & 0.5, 0.8 & 14,17,20 \\ [1ex]
      3 & \bf{superfine} & 2.896(6) & $48^3\times144$ & $250 \times 8$ & 0.0048 & 0.024 & 0.286 & 0.0234 & 
      0.274 & 0.427, 0.525, 0.65, 0.8  & 22,25,28  \\ [1ex]
      4 & \bf{ultrafine} & 3.892(12) &  $64^3\times192$ & $249 \times 4$ & 0.00316 & 0.0158 & 0.188 & 0.0165 
      & 0.194 & 0.5, 0.65, 0.8 & 31,36,41\\ [1ex]
      \hline
    \end{tabular}
  \end{center}
  \caption{Parameters for the ensembles of gluon field configurations that we 
use~\cite{Bazavov:2010ru,Bazavov:2012xda}. 
$a$ is the lattice spacing, determined from the Wilson flow parameter, $w_0$. Values 
for $w_0/a$ are from: set 1,~\cite{Chakraborty:2016mwy}, sets 2 and 3,~\cite{Chakraborty:2014aca} and set 4~\cite{craig}. The physical value of $w_0$ was determined at 0.1715(9) fm 
from $f_{\pi}$~\cite{Dowdall:2013rya}. $N_x$ is the spatial extent and $N_t$ the temporal extent of the lattice in lattice units; $n_{\text{cfg}}$ is the number of gluon field configurations 
in the ensemble and $n_{\text{src}}$ the number of different time sources used per 
configuration. 
Light, strange and charm quarks are included in the sea, their masses are given in 
columns 6-8, and the valence quark masses in columns 9-11. 
The $s$ and $c$ valence quarks were tuned in~\cite{Chakraborty:2014aca}. 
We use a number of heavy quark masses to assist the extrapolation to the physical $b$ 
mass. Column 12 gives the temporal separations between source and sink, $T$, of the 
3-point correlation functions computed on each ensemble. }
  \label{tab:ensembles}
\end{table*}

Staggered quarks have no spin degrees of freedom, so that solution of the 
Dirac equation on each gluon field is numerically fast. 
The remnant of the doubling problem means that 
quark bilinears of specific spin-parity have multiple copies, 
called `tastes'~\cite{Follana:2006rc}. 
They differ in the amount of point-splitting between the fields and the 
space-time dependent phase needed to substitute for the appropriate $\gamma$ matrix. 
In this calculation we can use only local (non point-split) bilinears, which is 
an advantage in terms of statistical noise, since no gluon fields are included in the 
current operator. In the standard staggered spin-taste notation, the 
 operators that we use are: pseudoscalar, $\Gamma_P = (\gamma^5\otimes \gamma^5)$; 
vector, $\Gamma_V^{\mu} = (\gamma^{\mu}\otimes \gamma^{\mu})$ and axial-vector, $\Gamma_A^{\mu} = (\gamma^{\mu}\gamma^5\otimes \gamma^{\mu}\gamma^5)$.

We compute several `two-point' correlation functions on the ensembles detailed 
in table \ref{tab:ensembles}, combining HISQ propagators from solving the Dirac 
equation for each random wall time source. 
These correlation functions take the form
\begin{align}
  C_{M}(t) =& \frac{1}{N_{\text{taste}}}\langle \Phi_M (t) \Phi_M^{\dagger}(0) \rangle, \\ 
  &\Phi_M(t) = \sum_{{\bf{x}}} \bar{q}({\bf{x}},t) \Gamma q'({\bf{x}},t), \nonumber
\end{align}
where $\langle \rangle$ represents a functional integral, $q,q'$ are valence 
quark fields of the flavours the $M$ meson is charged under, 
$\Gamma$ is the spin-taste structure of $M$ and $1/N_{\text{taste}}$ is the 
staggered quark normalisation for closed loops. The random-wall source and the 
sum over $\bf{x}$ at the sink project onto zero spatial momentum.  
We compute the correlation functions for all $t$ values, i.e. $0\leq t \leq N_t$.

The correlation function for the heavy-strange pseudoscalar meson, $H_s$, 
with valence quark content $h\overline{s}$ and spin-taste 
structure $\Gamma_P$ is constructed from HISQ propagators as:
\begin{align}
  C_{H_s}(t) = \frac{1}{4}\sum_{\bf{x},\bf{y}} \text{Tr}\left[ g_h(x,y) g_s^{\dagger}(x,y) \right].
  \label{eq:pseudoscalar_corrs}
\end{align}
Here $g_q(x,y)$ is a HISQ propagator for flavour $q$, 
the trace is over color and $1/4$ is the staggered quark normalisation. 
$x_0=0$ and $y_0=t$ and the sum is over spatial 
sites {$\bf{x}$, $\bf{y}$}. We also compute correlators for a 
charm-strange vector meson $D_s^*$, with structure $\Gamma^{i}_V$, using
\begin{align}
  C_{D_s^*}(t) = \frac{1}{4}\sum_{\bf{x},\bf{y}} (-1)^{x_{i}+y_{i}}\text{Tr}\left[ g_c(x,y) g_s^{\dagger}(x,y) \right].
\end{align}
We average over polarisations, $i=1,2,3$. 

We also compute correlation functions for two tastes of pseudoscalar heavy-charm mesons 
denoted $H_c$ and $\hat{H}_c$ respectively. $H_c$ has spin-taste 
structure $\Gamma_P$, and $\hat{H}_c$ has structure $\Gamma^0_A$. 
$H_c$ correlators are computed using Eq.~\eqref{eq:pseudoscalar_corrs} 
(with $g_s$ replaced with $g_c$), while $\hat{H}_c$ correlators are given by
\begin{align}
  C_{\hat{H}_c}(t) = \frac{1}{4}\sum_{\bf{x},\bf{y}}(-1)^{\bar{x}_0+\bar{y}_0} \text{Tr}\left[ g_h(x,y) g^{\dagger}_c(x,y) \right],
\end{align}
where we use the notation $\bar{z}_{\mu} = \sum_{\nu\neq\mu} z_{\nu}$.
These correlators will be used to normalise the axial vector $b\overline{c}$ current 
as discussed in Section~\ref{subsec:norm}.  

A useful physical proxy (that does not run) for the quark mass is that of 
the pseudoscalar meson made from that flavour of quark. 
It is therefore also useful, for our heavy quark mass extrapolation, to 
calculate correlation functions for heavy-heavy pseudoscalars, denoted $\eta_h$, with 
spin-taste structure $\Gamma_P$ using Eq.~\eqref{eq:pseudoscalar_corrs}. 
Likewise, to test the impact of any mistuning of the charm and 
strange quark masses, we also determine $\eta_c$ and $\eta_s$ correlators 
analogously. 
We can tune the $c$ and $b$ 
masses using the experimental values for the $\eta_c$ and $\eta_b$
masses, allowing for slight shifts from missing QED effects and the fact 
that we do not allow these mesons to annihilate to gluons~\cite{McNeile:2010ji}. 
The mass of the $\eta_s$ meson (which is not a physical state) can be fixed 
in lattice QCD from the $K$ and $\pi$ meson masses~\cite{Davies:2009tsa, Dowdall:2013rya}. 

We then generate the `three-point' correlation functions that contain 
the $H_s$ to $D_s^*$ transition. 
\begin{align}
  C_{\text{3pt}}(t,T) =& \frac{1}{N_{\text{taste}}}\sum_{{\bf{y}}} \langle \Phi_{D_s^*}(T)\, A^{i}({\bf{y}},t) \,\Phi_{H_s}(0) \rangle, \\
  &A^{\mu}({\bf{y}},t) = \bar{c}({\bf{y}},t) \gamma^5\gamma^{\mu} h({\bf{y}},t). \nonumber
\end{align}
Our $H_s$ source is given spin-taste $\Gamma_P$, the $D_s^*$ sink,  $\Gamma^{i}_V$, 
and the current insertion $\Gamma^{i}_A$. This gives the required cancellation of 
tastes
within the three-point function~\cite{Donald:2013pea}. 
In terms of HISQ propagators 
\begin{align}
  C_{\text{3pt}}(t,T) =& \frac{1}{4}\sum_{{\bf{x},\bf{y},\bf{z}}} (-1)^{\bar{y}_{i}+\bar{z}_{i}}  \nonumber \\ &\times\text{Tr}\left[ g_h(x,y)g_c(y,z) g^{\dagger}_s(z,x) \right],
\end{align}
where we fix $x_0 = 0$, $y_0=t$ and $z_0=T$. 
We compute the three-point correlation functions for all $t$ values 
within $0\leq t\leq T$, and 3 $T$ values that vary between 
ensembles and are given in Table~\ref{tab:ensembles}. We average over the 3 directions 
for $i$ for increased statistical precision. 

\subsection{Analysis of Correlation Functions}
\label{subsec:analysis}

We use simultaneous Bayesian fits~\cite{Lepage:2001ym, corrfitter} to extract the axial vector 
matrix element and meson masses from the two- and three-point correlation functions.
This allows us to include the covariance between results at different heavy quark 
masses on a given ensemble into our subsequent fits in Section~\ref{sec:extrapolation}. 

We fit the two-point correlation functions using the functional form
\begin{align}
\label{eq:2ptfit}
  C_M(t)|_{\text{fit}} &= \sum_{n}^{N_{\text{exp}}} \Big( | a^M_{n} |^2 f(E^M_n,t) \\ \nonumber
  &- (-1)^t | a^{M,o}_n |^2 f(E_n^{M,o},t) \Big); \\ \nonumber \\
  \nonumber
  f(E,t) &= \left( e^{-E t} + e^{-E (N_t-t)} \right),
\end{align}
where $N_t$ is the temporal extent of the lattice, and $E^{M,(o)}_n$,$a^{M,(o)}_n$ 
are fit parameters, with the excited-state energy parameters implemented as energy 
differences to the state below~\cite{Lepage:2001ym}. 
The second term accounts for the presence of opposite-parity 
states that contribute an oscillating term to the correlation function when 
using staggered quarks~\cite{Follana:2006rc}. 
These terms do not appear when $M$ is a pseudoscalar with a quark and 
antiquark of the same mass, so in the $M=\eta_h,\eta_c,$ and $\eta_s$ cases 
the second term is not required. For all correlator fits we set $N_{\text{exp}}=5$; 
this allows the impact of systematic effects from excited states to be 
included in the ground-state parameters that we are interested in.

The three-point correlation functions have the fit form
\begin{align}
  C_{\text{3pt}}&(t,T)|_{\text{fit}} = \sum_{k,j=0}^{N_{\text{exp}},N_{\text{exp}}} \\ 
  & \Big(\, a^{H_s}_j J^{nn}_{jk} a^{D_s^*}_k f(E^{H_s},t) f(E^{D_s^*}_n,T-t) \nonumber \\
  &+a^{H_s,o}_j J^{on}_{jk} a^{D_s^*}_k (-1)^t f(E^{H_s,o}_n,t) f(E^{D_s^*},T-t) \nonumber \\
  &+a^{H_s}_j J^{no}_{jk} a^{D_s^*,o}_k (-1)^{T-t} f(E^{H_s},t) f(E^{D_s^*,o}_n,T-t) \nonumber \\
  &+a^{H_s,o}_j J^{oo}_{jk} a^{D_s^*,o}_k (-1)^T f(E^{H_s,o}_n,t) f(E^{D_s^*,o},T-t) \,\Big). \nonumber
\end{align}
This includes fit parameters common to the fits of the $H_s$ 
and $D_s^*$ two-point correlators, along with new fit parameters $J_{jk}$. 

We perform a single simultaneous fit containing each correlator 
computed ($H_s,D_s^*,\eta_h,\eta_c,\eta_s,H_c,\hat{H}_c$, and three-point) for each ensemble.
We set gaussian priors for the parameters $J_{jk}$, and 
log-normal priors for all other parameters. Using log-normal distributions 
forbids energy differences $E_{n+1}^M - E_n^M$ and 
amplitudes $a_n^M$ (which can be taken to be positive here) 
from moving too
close to zero or changing sign, improving stability of the fit. 

Ground state energies $E_0^M$ are given priors of 
$(am_{q0} + am_{q'0} + a\Lambda_{\text{QCD}} )\pm 2a\Lambda_{\text{QCD}}$, 
where $m_{q0}$ and $m_{q'0}$ are the masses of the appropriate quarks, 
and $\Lambda_{\text{QCD}}$ is the confinement scale, which we set to 0.5GeV. 
For $q=h$ or $c$, this corresponds to the leading order HQET expression for 
a heavy meson mass. 
Ground-state energies of oscillating states, $E_0^{M,o}$, are given 
priors of $(am_{q0} + am_{q'0} + 2 a\Lambda_{\text{QCD}})\pm 2a\Lambda_{\text{QCD}}$. 
Excited state energy differences, $E_{i+1}^M - E_i^M$, $i>0$ are given prior values 
$2a\Lambda_{\text{QCD}}\pm a\Lambda_{\text{QCD}}$. 
Priors for ground state amplitudes $a_0^M$, are set from plots of 
effective amplitudes.  
The resulting priors always have a variance at least 10 times that of the 
final result for the ground-state. 
We use log(amplitude) priors of -1.20(67) for 
non-oscillating excited states and -3.0(2.0) for oscillating excited  
states. 
The ground-state non-oscillating to non-oscillating 3-point 
parameter, $J_{00}^{nn}$ is given a prior of $1\pm 0.6$, and the rest of 
the 3-point parameters $J_{jk}^{nn}$ are given $0\pm 1$.

$E_0^M = aM_M$ is the mass of the ground-state meson $M$ in lattice units. 
The masses $M_{H_s}$ and $M_{\eta_h}$ can both be used as proxys for $m_h$ 
in the extrapolation to $m_h=m_b$. 
The annihilation amplitude for an $M$-meson is given 
(in lattice units) by
\begin{align}
  \langle 0 | \Phi_M | M \rangle |_{\text{lat}} = \sqrt{2M_M} a_0^M .
  \label{eq:decayconstantfit}
\end{align}
The (as yet unnormalised) matrix element that 
we need to obtain $h_{A_1}^s(1)$ is given by
\begin{align}
  \langle D_s^*(\hat{k}) | A^k | H_s \rangle |_{\text{lat}} = 2 \sqrt{M_{H_s}M_{D_s^*}} J^{nn}_{00}.
  \label{eq:currentfit}
\end{align}

To ensure that truncating the sum over states at $N_{\text{exp}}$ accounts for the full 
systematic error from excited states, we cut out some data very 
close to the sources and sinks, where even higher excited states might have some effect. 
To do this we only include data with  $t \geq t_{\text{cut}}$ and $t \leq N_t-t_{\text{cut}}$ 
in the two-point case and $t \leq T-t_{\text{cut}}$ in the three-point case. 
We can in principle use a different $t_{\text{cut}}$ for every correlation 
function included in our fit, but we do not use a big range of $t_{\text{cut}}$ values. 
They range from 1 to 3 for the three-point functions and up to 8 for the two-point functions. 

The determination and minimisation of the $\chi^2$ function in our fit procedure 
requires the inversion of the covariance matrix that captures correlations between 
the different pieces of `data' (correlation functions) in our fit. The low 
eigenmodes of the correlation matrix are not well determined with the statistics 
that we have and so we implement an SVD (singular value decomposition) 
cut in the inversion of the correlation
matrix to avoid underestimating the uncertainty in the parameters of the 
fit~\cite{corrfitter}. 
This replaces correlation matrix eigenvalues below 
$\lambda_{\text{min}}$, equal to svdcut times the largest eigenvalue, with $\lambda_{\text{min}}$.  
$\lambda_{\text{min}}$ is estimated using the diagnosis tools 
in the Corrfitter package~\cite{corrfitter} 
and corresponds typically to an svdcut of $10^{-3}$ here.  

\begin{figure}[ht]
  \includegraphics[width=0.5\textwidth]{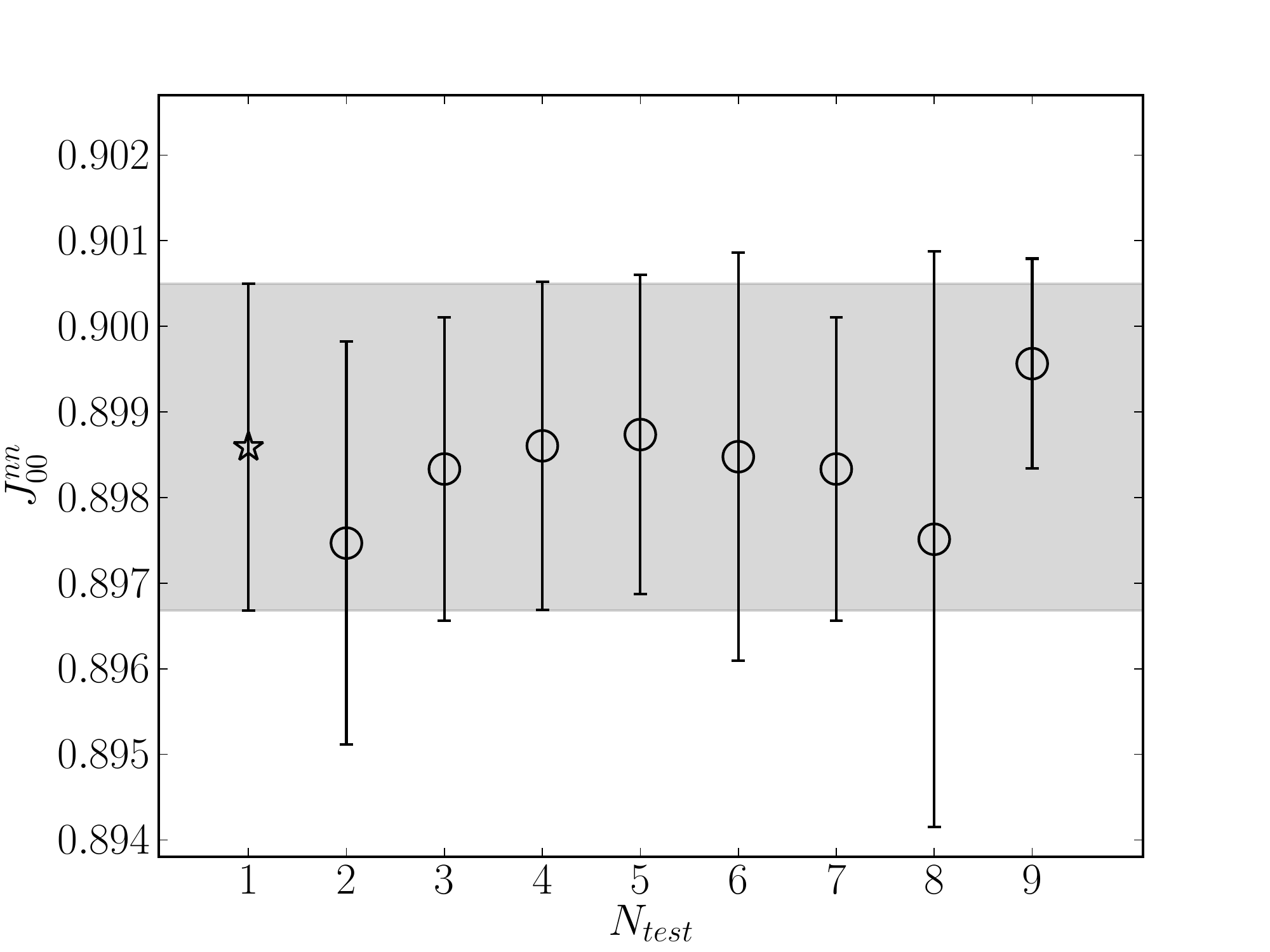}
  \caption{ 
Tests of the stability of correlator fits 
for $J_{00}^{nn}$ from fitting the two- and three-point correlators 
at heavy mass $am_{h0}^{\text{val}}=0.5$ on the fine ensemble. 
$N_{\text{test}}=1$ gives our final result. 
$N_{\text{test}}=2$ gives the results when all priors are broadened by 50\%. 
$N_{\text{test}}=3$ and $4$ give the results of setting 
$N_{\text{exp}}=4$ and $6$ respectively. 
$N_{\text{test}}=5, 6$ give the result of setting $t_{\text{cut}}=2,4$ 
respectively for all correlators. 
$N_{\text{test}}=7$ gives the result without marginalising out 
the $n=5$ excited state. 
$N_{\text{test}}=8$ gives the result of changing the SVD cut 
from $10^{-3}$ to $10^{-2}$.
    $N_{\text{test}}=9$ gives the result from a fit containing 
only $am^{\text{val}}_{h0}=0.5$ correlators and hence with a smaller covariance 
matrix. This allows us, as a test, to use 
a reduced SVD cut of $10^{-5}$. 
\label{fig:J00tests}}
\end{figure}

Figure~\ref{fig:J00tests} summarises stability tests of our fits, 
focussing on the key parameter $J_{00}^{nn}$ that is converted to 
the ground-state to ground-state transition amplitude using 
Eq.~(\ref{eq:currentfit}).

The fit parameters determined by our fits that we will use to 
calculate the physical value for $h^s_{A_1}(1)$ 
are given in Table~\ref{tab:results}. 
Notice that the statistical errors on the results grow with the 
heavy quark mass. This is a well understood problem in lattice heavy-light
meson physics (see, for example~\cite{Davies:2010ip}). 
Our method here has the advantage of including information from 
lighter-than-$b$ heavy quarks with improved statistical precision. 

\begin{table*}
\begin{center}
\begin{tabular}{ c c c c c c c c c c }
\hline
Set & $am_{h0}^{\text{val}}$ & $h^s_{A_1}(1)$ & $aM_{H_s}$& $aM_{D^*_s}$ & $aM_{H_c}$ & $af_{H_c}$ & $aM_{\eta_h}$ & $aM_{\eta_c}$ & $aM_{\eta_s}$ \\ [0.5ex]
\hline
1 & 0.5 & 0.9255(20) & 0.95972(12) & 0.96616(44) & 1.419515(41) & 0.186299(70) & 1.471675(38) & 1.367014(40) & 0.313886(75)\\ [0.5ex] 
 & 0.65 & 0.9321(22) & 1.12511(16) &  & 1.573302(40) & 0.197220(77) & 1.775155(34) & & \\ [0.5ex] 
 & 0.8 & 0.9434(24) & 1.28128(21) &  & 1.721226(39) & 0.207068(78) & 2.064153(30) & & \\ [0.5ex] 
\hline
2 & 0.5 & 0.9231(21) & 0.95462(12) & 0.93976(42) & 1.400034(28) & 0.183472(62) & 1.470095(25) & 1.329291(27) & 0.304826(52)\\ [0.5ex] 
 & 0.8 & 0.9402(27) & 1.27577(22) &  & 1.702456(23) & 0.203407(45) & 2.062957(19) & & \\ [0.5ex] 
\hline
3 & 0.427 & 0.9107(46) & 0.77453(24) & 0.63589(49) & 1.067224(46) & 0.126564(70) & 1.233585(41) & 0.896806(48) & 0.207073(96)\\ [0.5ex] 
 & 0.525 & 0.9165(49) & 0.88487(31) &  & 1.172556(46) & 0.130182(72) & 1.439515(37) & & \\ [0.5ex] 
 & 0.65 & 0.9246(65) & 1.02008(39) &  & 1.303144(46) & 0.133684(75) & 1.693895(33) & & \\ [0.5ex] 
 & 0.8 & 0.9394(66) & 1.17487(54) &  & 1.454205(46) & 0.137277(79) & 1.987540(30) & & \\ [0.5ex] 
\hline
4 & 0.5 & 0.9143(51) & 0.80245(24) & 0.47164(39) & 1.011660(32) & 0.098970(52) & 1.342639(65) & 0.666586(89) & 0.15412(17)\\ [0.5ex] 
 & 0.65 & 0.9273(62) & 0.96386(33) &  & 1.169761(34) & 0.100531(60) & 1.650180(56) & & \\ [0.5ex] 
 & 0.8 & 0.9422(72) & 1.11787(43) &  & 1.321647(37) & 0.101714(70) & 1.945698(48) & & \\ [0.5ex] 
\hline
\end{tabular}
    \caption{Values extracted from correlation function fits for $h^s_{A_1}(1)$, 
along with quantities required in our fits to determine a value at the 
physical point. Results are given on each gluon field ensemble for each valence heavy 
quark mass used. Results come from our simultaneous fits to two-point 
and three-point correlation functions: $h^s_{A_1}(1)$ values are determined 
using Eq.~(\ref{eq:normalizations}) and the ground-state meson masses in 
columns 4, 5, 6, 8, 9 and 10 from Eq.~(\ref{eq:2ptfit}).  
$f_{H_c}$ is the $H_c$ meson decay constant determined from Eq.~(\ref{eq:decayconstant}).  \label{tab:results} }
  \end{center}
\end{table*}

\subsection{Normalisation of the Axial Current}
\label{subsec:norm}

The partially-conserved axial-vector current for the HISQ action is a 
complicated linear combination of one-link and three-link lattice currents. 
In this study we use only local axial vector currents. This simplifies 
the lattice QCD calculation but creates the need for our resulting current 
matrix element to be multiplied by a matching factor $Z_A$ to produce the 
appropriate continuum current. We determine $Z_A$ via a fully non-perturbative 
method~\cite{Donald:2013pea}.

We use the fact that the staggered local pseudoscalar 
current of spin-taste $(\gamma^5\otimes \gamma^5)$, multiplied by 
the sum of its valence quark masses, is absolutely normalized via the PCAC relation. 
From the two-point $H_c$ and $\hat{H}_c$ correlator fits we can extract 
the decay amplitudes: $\langle 0 | \bar{c} (\gamma^5\otimes \gamma^5) h | H_c \rangle \equiv \langle 0 | P | H_c \rangle$ 
and $\langle 0 | \bar{c} (\gamma^0\gamma^5 \otimes \gamma^0\gamma^5) h | \hat{H}_c \rangle = \langle 0 | A^0 | \hat{H}_c \rangle$ as in Eq.~\eqref{eq:decayconstantfit}. 
Then, the normalization for the local $A^0$ current 
(common to that of the local 
spatial axial-vector current $A^k$ up to discretisation effects), 
$Z_A$, is fixed by demanding that 
\begin{align}
  (m^{\text{val}}_{h0} + m^{\text{val}}_{c0}) \langle 0 | P | H_c \rangle|_{\text{lat}} = M_{\hat{H}_c} Z_A \langle 0 | A^0 | \hat{H}_c \rangle|_{\text{lat}}.
  \label{eq:ward}
\end{align}
The $Z_A$ values found on each ensemble and $am^{\text{val}}_{h0}$ are given in Table~\ref{tab:norms}.

There is an ambiguity in what mass to use on the right hand side of Eq.~\eqref{eq:ward}. 
We use the non-goldstone mass $M_{\hat{H}_c}$, but one could just as well replace 
this with $M_{H_c}$ since the difference is a discretisation effect. 
The meson mass difference is very small for heavy mesons~\cite{Follana:2006rc} 
and so we find the effect of changing the taste of meson mass used 
never exceeds 0.15\% of $Z_A$ throughout the range of ensembles and heavy 
masses that we use and has no impact on the continuum result. 

We also remove tree-level mass-dependent discretisation effects coming from 
the wavefunction renormalisation~\cite{Follana:2006rc} by multiplying by 
a factor $Z_{\text{disc}}$. This is derived in~\cite{Monahan:2012dq}  
as:
\begin{align}
  \label{eq:Zdisc}
  Z_{\text{disc}} &=\,\sqrt{ \tilde{C}_h \tilde{C}_c }, \\
  &\tilde{C}_q = \cosh am_{q,\text{tree}} \left( 1 - {1+\epsilon_{q,\text{Naik}}\over 2} \sinh^2 am_{q,\text{tree}} \right).  \nonumber
\end{align}
See also~\cite{Bazavov:2017lyh}. 
$m_{q,\text{tree}}$ is the tree-level pole mass in the HISQ action. It has an expansion in terms 
of the bare mass~\cite{Follana:2006rc}
\begin{align}
\label{eq:mexp}
  &am_{q,\text{tree}} = am_{q0} \Big( 1 - {3\over 80}am_{q0}^4 + {23\over 2240}am_{q0}^6 \\ \nonumber
  &+ {1783\over 537600} am_{q0}^8 - {76943\over23654400}am_{q0}^{10} + \order{am_{q0}^{12}} \Big), \nonumber
\end{align}
$\epsilon_{q,\text{Naik}}$ fixes the Naik parameter~\cite{Naik:1986bn} 
($N=1+\epsilon$ is the coefficient of the 
tree-level improvement term for the derivative) in the HISQ action when it is being 
used for heavy quarks~\cite{Follana:2006rc}. 
$\epsilon_{q,\text{Naik}}$ is set 
to its tree-level value, removing the leading tree-level errors 
from the dispersion relation. As an expansion in $am_{q,\text{tree}}$ it begins at 
$\mathcal{O}(am_{q,\text{tree}})^2$~\cite{Follana:2006rc}. 
To determine $\epsilon_{q, \text{Naik}}$ we use the 
 closed form expression for it given 
in~\cite{Monahan:2012dq} and this can also be 
used along with Eq.~(\ref{eq:mexp}) to evaluate 
$Z_{\text{disc}}$. 
The pole condition can be used to show 
that the expansion of $\tilde{C}_q$ begins at $am^4_{q0}$ as 
$1-3am^4_{q0}/80 + \ldots$.   
The effect of $Z_{\text{disc}}$ is then very small, never exceeding $0.2\%$. 
$Z_{\text{disc}}$ values on each ensemble for each $am^{\text{val}}_{h0}$ are given 
in table \ref{tab:norms}.

Combining these normalizations with the lattice current from the 3-point fits, 
we find a value for the form factor at a given heavy mass and lattice spacing:
\begin{align}
  h_{A_1}^s(1) = {1\over 3} \sum_{k=1}^3 { Z_A Z_{\text{disc}}\langle D_s^*(\hat{k}) | A^k | H_s \rangle |_{\text{lat}}\over 2\sqrt{M_{H_s} M_{D_s^*}} }.
  \label{eq:normalizations}
\end{align}

\begin{table}
\begin{center}
\begin{tabular}{ c c c c }
\hline
Set & $am_{h0}^{\text{val}}$ & $Z_A$& $Z_{\text{disc}}$\\ [0.5ex]
\hline
1 & 0.5 & 1.03178(57) & 0.99819\\ [0.5ex] 
 & 0.65 & 1.03740(58) & 0.99635\\ [0.5ex] 
 & 0.8 & 1.04368(56) & 0.99305\\ [0.5ex] 
\hline
2 & 0.5 & 1.03184(47) & 0.99829\\ [0.5ex] 
 & 0.8 & 1.04390(39) & 0.99315\\ [0.5ex] 
\hline
3 & 0.427 & 1.0141(12) & 0.99931\\ [0.5ex] 
 & 0.525 & 1.0172(12) & 0.99859\\ [0.5ex] 
 & 0.65 & 1.0214(12) & 0.99697\\ [0.5ex] 
 & 0.8 & 1.0275(12) & 0.99367\\ [0.5ex] 
\hline
4 & 0.5 & 1.00896(44) & 0.99889\\ [0.5ex] 
 & 0.65 & 1.01363(49) & 0.99704\\ [0.5ex] 
 & 0.8 & 1.01968(55) & 0.99375\\ [0.5ex] 
\hline
\end{tabular}
    \caption{Normalization constants applied to the lattice axial vector current in \eqref{eq:normalizations}. $Z_A$ is found from \eqref{eq:ward} and $Z_{\text{disc}}$ from \eqref{eq:Zdisc}. \label{tab:norms}}
  \end{center}
\end{table}

\subsection{Obtaining a Result at the Physical Point}
\label{sec:extrapolation}

We now discuss how we fit our results for the zero recoil form factor, $h_{A_1}^s(1)$, 
as a function of valence heavy quark mass, sea light quark mass and 
lattice spacing to obtain a result at the physical point where the heavy quark mass 
is that of the $b$, the sea quark masses are physical and the lattice spacing is zero. 

In summary, we fit our results for $h_{A_1}^s(1)$ to the following form 
\begin{eqnarray}
\label{eq:fitform}
  h^s_{A_1}(1)(a,m_l,m_h) &=  1 - \left({\varepsilon_c\over 2}\right)^2 l_V + {\varepsilon_c\varepsilon_h} \frac{l_A}{2} - \left({ \varepsilon_h \over 2 }\right)^2 l_P \nonumber \\ 
&+ \mathcal{N}_{\text{disc}} + \mathcal{N}_{\text{mistuning}}.
\end{eqnarray}
The terms in the first line allow for dependence on the valence heavy quark and 
charm quark masses  (with $\varepsilon_q \equiv 1/m_q$)
using input from HQET, to be discussed below.
$\mathcal{N}_{\text{disc}}$ and $\mathcal{N}_{\text{mistuning}}$ account for 
discretisation and mass mistuning effects, also discussed below. 
The physical result is then $h^s_{A_1}(1)(0,m_{l,\text{phys}},m_b)$. 

\subsubsection{Dependence on the heavy valence quark mass}
\label{sec:heavymass}

Our fit of the $m_h$ dependence is guided by HQET, which considers both the 
$c$ quark and the heavy quark of mass $m_h$ to be heavy here. 
In particular, for the 
parameter $h^{(s)}_{A_1}(1)$, HQET forbids terms of $\mathcal{O}(1/m_Q)$ where 
$m_Q$ can be $m_c$ or $m_b$~\cite{Luke:1990eg}.  
The HQET expression for $h_{A_1}(1)$ is then given by~\cite{Falk:1992wt,Mannel:1994kv}:
\begin{align}
  \label{eq:hqet}
  h_{A_1}(1) &= \eta_A \left( 1 - {l_V\over (2m_c)^2} + {l_A \over 2 m_c m_h} - { l_P \over (2m_h)^2 } \right)  \\ \nonumber &+ \mathcal{O}\left( \, {1\over m_c^n m_h^m}, \, n+m\geq 3 \, \right),
\end{align}
where $l_V$, $l_A$ and $l_P$ are $\mathcal{O}(\Lambda^2_{\text{QCD}})$ (with possible 
mild dependence on whether the spectator quark is $s$ or $u/d$). 
$\eta_A$ accounts for ultraviolet matching between HQET and QCD, 
and has been computed to 2-loops in perturbative QCD~\cite{Czarnecki:1996gu}. 
It has mild dependence on $m_h$ through logarithms of $m_c/m_h$; at one-loop $\eta_A$ has 
explicit form~\cite{Close:1984ad} 
\begin{equation}
\label{eq:eta-pert}
\eta_A =  1- \frac{\alpha_s}{\pi}\left(\frac{1+m_c/m_h}{1-m_c/m_h}\ln\frac{m_c}{m_h} + \frac{8}{3}\right) .
\end{equation}
The coefficient of $\alpha_s/\pi$ then varies from -0.66 to -0.29 across the range of 
$m_h$ from $m_h=m_c$ to $m_h=m_b$, taking $m_b/m_c = 4.577(8)$~\cite{Bazavov:2018omf}. 
The two-loop correction 
is small~\cite{Czarnecki:1996gu}. $\eta_A$ is then 
close to 1 and differs by a few percent across our range of $m_h$. 

Our calculation has results at multiple values of $m_h$, and 
could therefore in principle provide 
information on the coefficients $l_A$ and $l_P$ 
of the $m_h$-dependent terms in the HQET expansion.  
The charm quark mass is fixed to its physical value and so we cannot access
the value of $l_V$ independent of a choice of $\eta_A$ at $m_h=m_c$. 
The terms in round brackets in Eq.~(\ref{eq:hqet}), multiplying $\eta_A$, are all 
very small because of the suppression by heavy-quark 
masses. To constrain them tightly requires very precise data and, as we will 
see, we are not able to determine $l_A$, $l_P$ or $l_V$ accurately with our 
results. It therefore does not make sense to attempt to compare them accurately to 
HQET expectations. To do so would require using an appropriate quark mass 
definition (since different definitions will move quark mass dependence between 
the $l_A$ term and the others in Eq.~(\ref{eq:hqet}) ) and the two-loop expression 
for $\eta_A$ with appropriate value for $\alpha_s$ (since logarithmic $m_h$ dependence of 
$\eta_A$ can be misinterpreted as part of a polynomial in $1/m_h$). 

Instead we simply take an HQET-inspired form for the $m_h$-dependence 
and set $\eta_A$ to 1, resulting in the first line of our fit form, 
Eq.~(\ref{eq:fitform}). 
This is sufficient to test, 
through the results we obtain for $l_A$, $l_V$ and $l_P$ using this expression, 
that the HQET expectation for the approximate size of these coefficients is fulfilled. 
We take priors on $l_{A,V,P}$ in our fit of $0\pm 1 \,\text{GeV}^2$.  

We have several different proxies, derived from heavy meson masses, 
that we can take for the heavy quark 
mass that appears in $\varepsilon_h$ in Eq.~(\ref{eq:fitform}). We do not 
expect our physical result for $h^s_{A_1}$ to vary significantly depending on 
which meson mass we use, but the results for $l_A$, $l_V$ and $l_P$ will vary because 
of different sub-leading terms in the relationship between meson and quark mass. 
The most obvious substitutions to use for the heavy quark mass are the mass 
of the pseudoscalar heavy-strange 
meson, $M_{H_s}$, and half the mass of the pseudoscalar heavyonium meson, $M_{\eta_h}$.
We also tested using the quark mass in the minimal renormalon subtracted (MRS) scheme 
suggested in~\cite{Brambilla:2017hcq}. This takes 
\begin{align}
  \label{eq:hqet_mass}
  m_h = M_{H_s} - \bar{\Lambda}_\text{MRS} - { \mu^2_\text{MRS} \over M_{H_s} - \bar{\Lambda}_\text{MRS}} + \mathcal{O}\left({1\over m_h^2}\right).
\end{align}
where 
$\mu_\text{MRS}^2 = \mu^2_{\pi,\text{MRS}} - d_{H^{(*)}} \mu^2_{G,\text{MRS}}$ with $d_{H^{(*)}} = 1$ 
for pseudoscalar mesons and $-1/3$ for vectors. 
For this case we use parameters determined in~\cite{Bazavov:2018omf} for the MRS scheme: 
$\bar{\Lambda}_{\text{MRS}} = 0.552(30)\text{GeV}$, 
$\mu^2_{\pi,\text{MRS}} = 0.06(22)\text{GeV}^2$ and 
$\mu^2_{G,\text{MRS}} = 0.38(1)\text{GeV}^2$. We take $m_h$ from 
Eq.~(\ref{eq:hqet_mass}) using our results for the mass of  
the pseudoscalar heavy-strange meson and $m_c$ from our results for the mass of the 
$D_s^*$ meson. 

We take our central fit, for simplicity, from the result of using 
half the pseudoscalar heavyonium 
mass for $m_h$ and half the pseudoscalar charmonium mass for $m_c$ i.e. taking 
\begin{equation}
\label{eq:epsdef}
\varepsilon_q \equiv \frac{2}{M_{\eta_q}} .
\end{equation}
We test the stability of the 
fit results under the different choices discussed 
above in Section~\ref{sec:stability}. 
 
\subsubsection{Mistuning of other quark masses}

Our calculation has results for multiple different heavy quark masses on each 
gluon field configuration. The valence charm and strange quark masses, however, 
are tuned to their physical values. This is done by fixing the $\eta_c$ and 
$\eta_s$ meson masses to their physical values in a pure QCD world allowing, for 
example, for $\eta_c$ annihilation as 
discussed in~\cite{Chakraborty:2014aca}.  
Any possible mistuning of the charm quark mass is accounted for in our fit 
function by the dependence on the charm quark mass that is included in the 
first line of Eq.~({\ref{eq:fitform}}). When the fit function is evaluated at the 
physical point we set $\varepsilon_c$ from the physical $\eta_c$ mass. 

The strange (valence and sea) and light (sea) mass mistunings are accounted 
for using the tuning in \cite{Chakraborty:2014aca}. 
For the strange quark, we define $\delta_s = m_s - m_s^{\text{tuned}}$, 
where $m_s^{\text{tuned}}$ is given by
\begin{align}
  m_s^{\text{tuned}} = m_{s0} \left({ M^{\text{physical}}_{\eta_s} \over M_{\eta_s} }\right)^2.
\end{align}
$M_{\eta_s}^{\text{physical}}$ is determined in lattice simulations 
from the masses of the pion and kaon~\cite{Dowdall:2013rya}.
The ratio $\delta_s/m_s^{\text{tuned}}$ then gives the fractional mistuning. 
The valence strange quark masses are very well tuned, but the sea strange 
quark masses less so. 

We similarly account for mistuning of the masses of the (sea) light quarks 
by defining $\delta_l = m_{l0} - m_l^{\text{tuned}}$. 
We find $m_l^{\text{tuned}}$ from $m_s^{\text{tuned}}$, 
leveraging the fact that the ratio of quark masses is regularization 
independent, and the ratio was calculated in \cite{Bazavov:2017lyh}:
\begin{align}
  \left.\frac{m_s}{m_l}\right\rvert_{\textrm{phys}} = 27.18(10).
\end{align}
We set $m_l^{\text{tuned}}$ to $m_s^{\text{tuned}}$ divided by this ratio.

The full term we include to account for mistuning is then given by
\begin{align}
  \mathcal{N}_{\text{mistuning}} = \frac{c^{\text{val}}_s {\delta^{\text{val}}_s} + c_s {\delta_s} + 2 c_l \delta_l}{10 m_s^{\text{tuned}}} 
  \label{eq:mistuning}
\end{align}
where $c_l$, $c_s$ and $c_s^{\text{val}}$ 
are fit parameters with prior distributions $0\pm 1$. 
We neglect $\delta_{s,l}^2$ contributions since these are an order of magnitude 
smaller and are not resolved in the results of our lattice calculation.

The gluon field configurations that we use have $m_u=m_d\equiv m_l$ in the sea. 
In the real world this is not true. 
We test the impact of possible isospin-breaking on our fits by testing for sensitivity 
to the sea light quark masses. We do this by changing the $m_l^{\text{tuned}}$ 
value up and down by the expected value for $m_d-m_u$~\cite{Tanabashi:2018oca}. 
We find the effect to be completely negligible in comparison to 
the other sources of error.

\subsubsection{Discretisation Effects}
\label{subsubsec:disc}

Discretisation effects in our lattice QCD results are accounted for following the methodology of~\cite{McNeile:2012qf}. We take
\begin{align}
    \label{eq:fitfun_hA1}
  \mathcal{N}_{\text{disc}} = &\sum_{i=0,j+k\neq 0}^{2,2,2} d_{ijk} \left({2\Lambda_{\text{QCD}}\over M_{\eta_h} }\right)^{i} \left({ am^{\text{val}}_{h0} \over \pi }\right)^{2j} \left({ am^{\text{val}}_{c0} \over \pi }\right)^{2k}.
\end{align}
The leading terms, with $i=0$, allow for discretisation effects that are set by 
the heavy quark mass and also discretisation effects that are set by the 
charm quark mass (or indeed any other lighter scale that is independent 
of heavy quark mass). 
The $i>0$ terms allow for discretisation effects to vary as 
the heavy quark mass is varied, 
with $M_{\eta_h}$ being used here as a proxy for the heavy quark mass. 
$d_{ijk}$ are fit parameters with prior distributions $0\pm 1.0$. 
All discretisation effects are of even order by construction 
of the HISQ action~\cite{Follana:2006rc}.

We tested the impact on the fit of including extra discretisation effects 
set by the scale $\lqcd$ but this made no difference (since such effects are much 
smaller than those already included by the $am^{\text{val}}_{c0}$ terms). 
We also tested the effects of increasing the number of terms 
in each sum, but the final result remained unchanged.

\subsubsection{Finite-volume Effects}

The finite volume effects in our lattice results are expected to be negligible, 
because we are working with heavy mesons that have no valence light quarks 
and no Zweig-allowed strong decay modes. Coupling to chiral loops or decay channels 
with pions that could produce significant finite-volume 
effects~\cite{Laiho:2005ue} is therefore
absent and we can safely ignore finite-volume effects here.  

In section \ref{sec:stability} we detail the results of several tests of the 
stability of our final result under changes to the details of the fit.

    \subsubsection{Topological Charge Effects}

    It has been observed that the finest MILC 
ensembles ($a\simeq 0.45$fm and finer) suffer from slow 
variation in the topological charge~\cite{Bernard:2017npd} with Monte Carlo 
time. 
The question then arises if physical observables obtained by averaging over 
the ensemble could be biassed by being measured in only a small 
range of topological 
charge sectors. 
This issue is addressed in~\cite{Bernard:2017npd} through a calculation of
the `topological adjustment' needed for meson masses and decay constants 
on the ultrafine lattices used here (with $m_l/m_s=0.2$). 
The adjustment found for the $D_s$ decay constant is 0.002\%. 
We might expect the impact of a frozen topological charge on 
$h_{A_1}^s(1)$ to be of a similar size to this, given that it involves a 
transition between heavy-strange mesons. 
Allowing for this systematic uncertainty (or even ten times it) 
has a negligible 
effect on our final result.  

\section{Results and Discussion}
\label{sec:results}

\subsection{Result for $h^s_{A_1}(1)$}

The results of our correlation function fits (discussed 
in Section~\ref{subsec:analysis}) are given in Table~\ref{tab:results}. 
We tabulate values for $h^s_{A_1}(1)$ at each heavy quark mass that we 
have used on each gluon field ensemble from Table~\ref{tab:ensembles}.  
We also tabulate the meson masses needed to allow determination of $h^s_{A_1}(1)$ 
at the physical point, using the fit form of Eq.~\ref{eq:fitform}. 

\begin{figure}[ht]
  \includegraphics[width=0.5\textwidth]{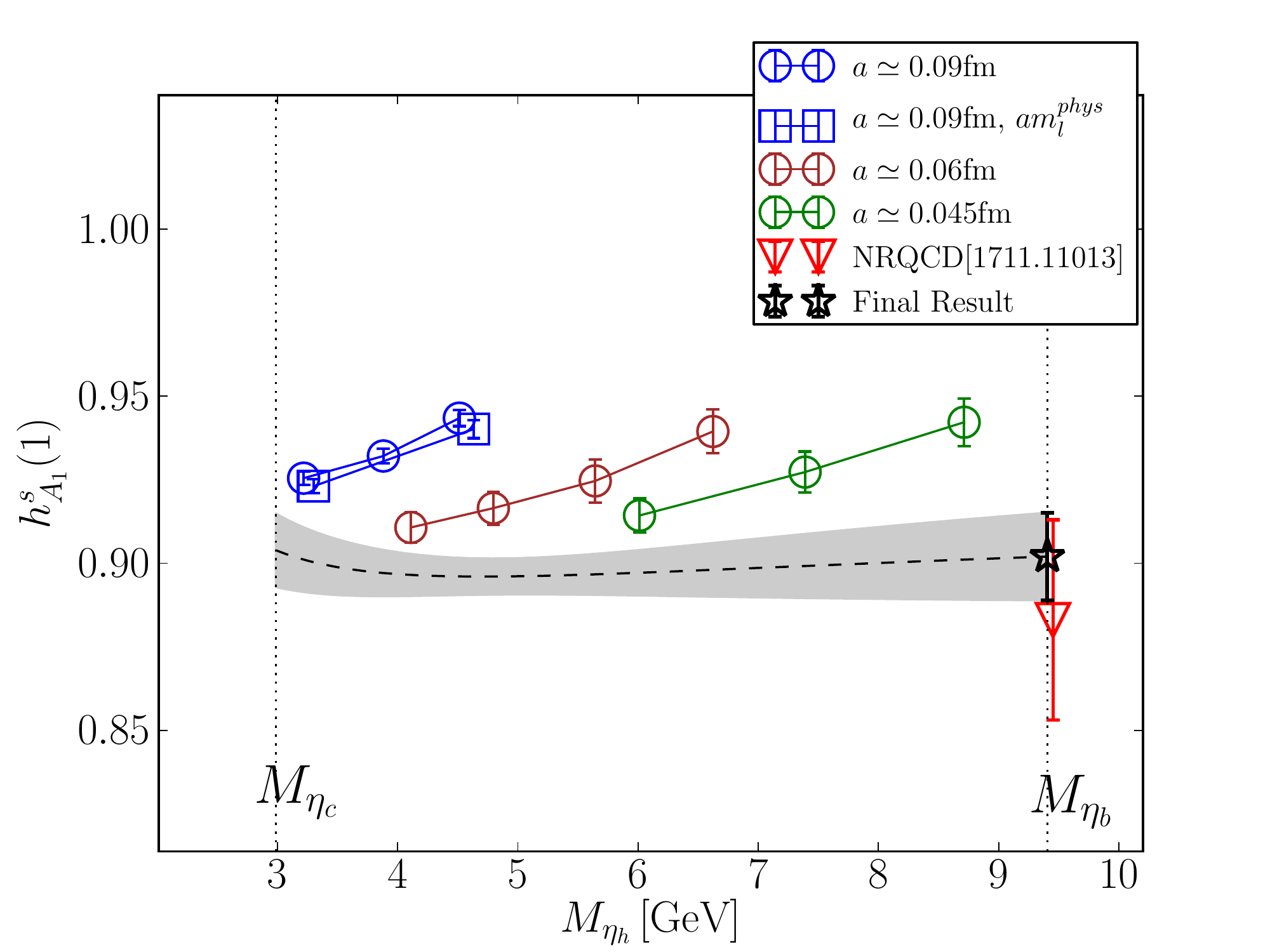}
  \caption{$h_{A_1}^s(1)$ against $M_{\eta_h}$ (a proxy for 
the heavy quark mass). The grey band shows the result of the extrapolation to $a=0$ 
at physical $l$,$s$ and $c$ masses; the black star shows our result at the 
physical $b$ quark mass. Gluon field ensembles listed in the 
legend follow the order of sets in Table~\ref{tab:ensembles}. Solid lines simply 
join the points on a given ensemble for added clarity.  
The red inverted triangle gives the determination of the same quantity from a 
previous study using the NRQCD action for the $b$ 
quark~\cite{Harrison:2017fmw}. \label{fig:hA1_vsmetah}}
\end{figure}

The fit function of Eq.~(\ref{eq:fitform}) is readily applied, giving a
$\chi^2/[\text{dof}]$ of 0.21 for 12 degrees of freedom. 
Figure~\ref{fig:hA1_vsmetah} shows our results for $h^s_{A_1}(1)$ along with 
the fit function at zero lattice spacing and physical $u/d$, $s$ and $c$ 
quark masses as the grey band. 
Evaluating the fit function at the physical $b$ mass, as determined by 
$M_{\eta_b}$, gives our final result 
\begin{align}
  \mathcal{F}^{B_s\to D_s^*}(1) = h^s_{A_1}(1) = 0.9020(96)_{\text{stat}}(90)_{\text{sys}}\,.
  \label{eq:finalresult}
\end{align}
Adding the statistical and systematic errors in quadrature, we find a 
total fractional error of $1.5\%$. 
The error budget for this result is given in table~\ref{tab:errorbudget}.
Note that we allow for an additional $\pm 10 \,{\text{MeV}}$ uncertainty 
in the physical value of the $\eta_b$ 
mass beyond the experimental uncertainty, since our lattice QCD results do not include 
the effect of $\eta_b$ annihilation and QED~\cite{McNeile:2012qf}. 
This has no effect, however, since 
the heavy quark mass dependence is so mild. 

\begin{table}
\begin{center}
    \begin{tabular}{c c}
      \hline
      Source & \% Fractional Error \\ [0.5ex]
      \hline
      Statistics $+Z_A$ & 1.06  \\ [1ex]
      $a\to 0$ & 0.73  \\ [1ex]
      $m_h \to m_b$ & 0.69  \\ [1ex]
      mass mistuning & 0.20  \\ [1ex]
      \hline
      Total & 1.45 \\ [1ex]
      \hline
    \end{tabular}
  \end{center}
  \caption{Error budget for $h^s_{A_1}(1)$. Errors are given as a percentage of 
the final answer. The mass mistuning error includes that from valence strange 
and sea light and strange quarks; we find that taking 
a $\pm 10 \,\text{MeV}$ uncertainty in the 
physical value of the $\eta_b$ mass has a negligible effect. \label{tab:errorbudget}}
\end{table}

Our total uncertainty is dominated by the statistical errors in our lattice results. 
The systematic error is dominated by that from the continuum extrapolation. 

We include in Figure~\ref{fig:hA1_vsmetah} the value from the only other 
lattice determination of $h^s_{A_1}(1)$~\cite{Harrison:2017fmw}. 
This calculation also used MILC $n_f=2+1+1$ gluon field ensembles, but 
with the bulk of the ensembles used having coarser lattice spacing. This was 
made possible by the use of the NRQCD action for the $b$ quark~\cite{Dowdall:2011wh}. 
The HISQ action was used for all the other quarks. 
The result of this calculation was: 
$h_{A_1}^s(1) = 0.883(12)_{\text{stat}}(28)_{\text{sys}}$. 
Our result is in agreement with this, but with substantially smaller errors. 
The NRQCD uncertainty of 3.4\% is dominated by the systematic error from 
the $\mathcal{O}(\alpha_s)$ matching factor used to normalise
the NRQCD-HISQ current and this error is absent from our calculation. 

In addition to a value for $h^s_{A_1}(1)$ 
our calculation is able to give information on the physical 
dependence on the heavy quark mass of $\mathcal{F}^{H_s \rightarrow D^*_s}(1)$. 
We see from Figure~\ref{fig:hA1_vsmetah} that this dependence is very 
mild to the point of being absent. We can determine the ratio 
of $\mathcal{F}^{H_s \rightarrow D^*_s}(1)$ for $m_h=m_b$ to $m_h=m_c$ (albeit 
that this latter point corresponds to an unphysical $D_s \rightarrow D_s^*$ 
decay) and find the value 0.998(23). Each of the terms (including $\eta_A$) 
in the HQET expectation 
of Eq.~\ref{eq:hqet} can give effects of order a few percent to this 
ratio. The fact that we find no heavy quark mass dependence at the level of 
2\% shows that these effects must tend to cancel out.   

The fit of our lattice results to Eq.~(\ref{eq:fitform}) gives 
fit parameters $l_{V,A,P}$ which, as discussed in 
Section~ \ref{sec:heavymass}, provide a test of HQET. We find 
\begin{align}
  \nonumber  l^s_V &= 0.71(28)\text{GeV}^2, \\  l^s_A &= -0.34(32)\text{GeV}^2, \label{eq:hqet_constants}
  \\ \nonumber l^s_P &= -0.53(34)\text{GeV}^2,
\end{align}
from our baseline fit. These results are compatible with values of 
$\mathcal{O}(\Lambda^2_{\text{QCD}})$ as expected by HQET. 
As discussed in Section~\ref{sec:heavymass} these fit parameters 
change depending on the proxy that we use for the quark mass as well as our treatment 
of $\eta_A$. However, as we show in the tests performed in the next section 
(see Figure~\ref{fig:fittests}) this has little impact on our value for 
$h^s_{A_1}(1)$. 

\subsection{Further tests of our fit}
\label{sec:stability}

\begin{figure}[ht]
  \includegraphics[width=0.5\textwidth]{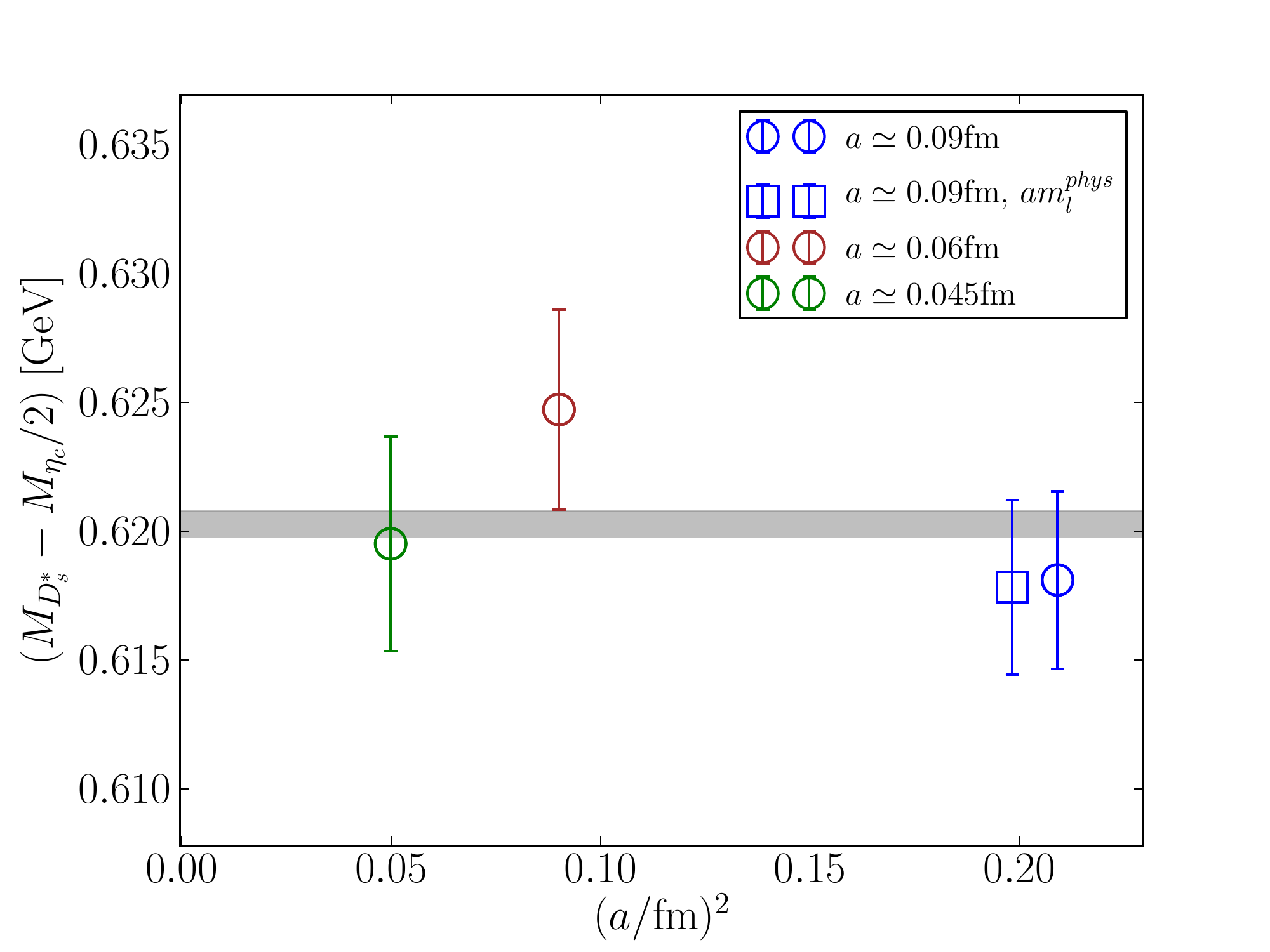}
  \caption{ The $D_s^*$ meson mass obtained on each of our gluon field ensembles, 
given as a difference from one half the $\eta_c$ meson mass. Errors include statistical 
and lattice spacing uncertainties. 
The grey band gives the experimental result~\cite{Tanabashi:2018oca}. 
 \label{fig:Dsmass}}
\end{figure}

\begin{figure}[ht]
  \includegraphics[width=0.5\textwidth]{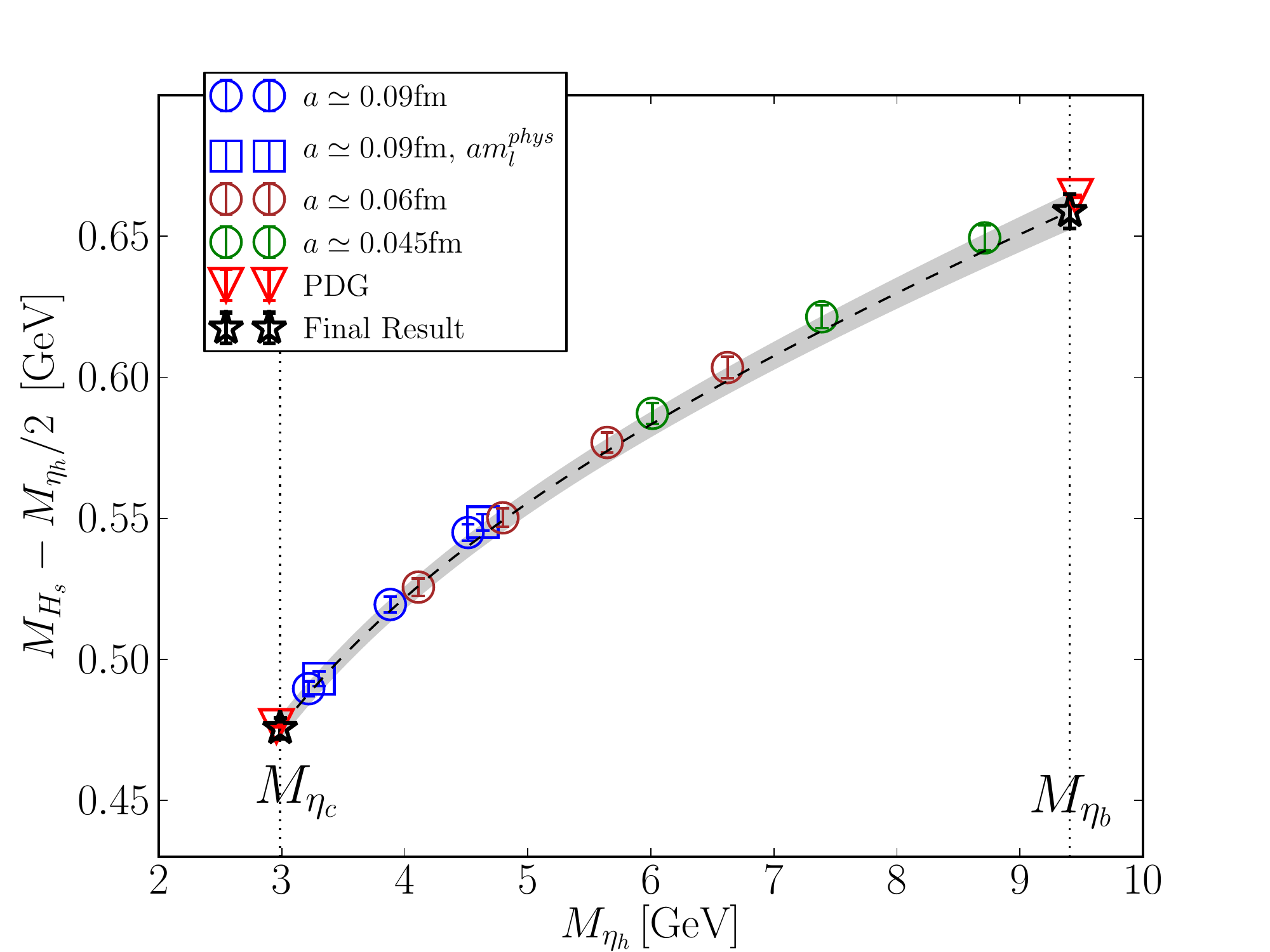}
  \caption{ The $H_s$ meson mass obtained on each of our gluon field ensembles, 
given as a difference to one half of the $\eta_h$ meson mass. Errors include statistical 
and lattice spacing uncertainties. 
The grey band gives a fit to the heavy-quark mass dependence as discussed 
in the text, with black stars giving our results at $m_h=m_c$ and $m_h=m_b$. 
The inverted red triangles give the corresponding experimental 
values~\cite{Tanabashi:2018oca}.
 \label{fig:MHs}}
\end{figure}

Because we tune our $b$ and $c$ valence quark masses 
using the pseudoscalar heavyonium meson mass, we can independently 
test 
our results by comparing both our heavy-strange and 
$D_s^*$ meson masses against experiment. 
These results are shown in Figures~\ref{fig:Dsmass} and~\ref{fig:MHs}. 
In each case we subtract half the corresponding pseudoscalar heavyonium 
mass to reduce lattice spacing uncertainties in the comparison to 
experiment~\cite{Davies:2010ip}. 

Figure~\ref{fig:Dsmass} shows that our $D_s^*$ meson mass 
agrees with experiment on all our ensembles at the level of our 5 MeV 
uncertainties. Systematic effects from missing QED and $\eta_c$ annihilation 
are expected to be of size a few MeV~\cite{Davies:2010ip}.  

Figure~\ref{fig:MHs} shows our results for the heavy-strange pseudoscalar meson 
mass as a function of the pseudoscalar heavyonium mass. We show the difference 
$\Delta_h = M_{H_s}-M_{\eta_h}/2$ to remove the 
leading $m_h$ dependence and also to reduce 
uncertainties from the value of the lattice spacing.  We fit $\Delta_h$ to a simple 
function of $\varepsilon_h$ (Eq.~(\ref{eq:epsdef})) :
\begin{eqnarray}
\Delta_h(a,m_l,m_h) &=& (\sum_{i=-1}^{i=1} c_i\varepsilon_h^i) \\
&\times& (1 + \mathcal{N}_{\text{disc}} + \mathcal{N}_{\text{mistuning}}) . \nonumber
\end{eqnarray}
The leading, linear, term in $\varepsilon_h$ allows for the fact that 
the heavyonium ($\eta_h$) binding energy grows linearly with $m_h$ in a $1/r$ potential. 
We take priors on the $c_i$ of: $c_{-1}:0.05(5)$; $c_0:0.5(5)$; $c_1:0(1)$. 
$\mathcal{N}_{\text{disc}}$ takes the same form as in Eq.~(\ref{eq:fitfun_hA1})
with $a\Lambda_{\text{QCD}}$ (where $\Lambda_{\text{QCD}}$ is taken as 0.5 GeV) replacing $am_{c0}^{\text{val}}$, 
which is not relevant here. $\mathcal{N}_{\text{mistuning}}$ takes the same 
form as in Eq.~(\ref{eq:mistuning}). 
 
Our result for the difference $M_{H_s}-M_{\eta_h}/2$ in the continuum 
at $m_h=m_c$ is 0.4755(37) GeV and at $m_h=m_b$ is 0.6588(61) GeV. 
These agree well with the earlier HPQCD results on 
$n_f=2+1$ gluon field configurations of 0.4753(22) GeV~\cite{Davies:2010ip} and 
0.658(11) GeV~\cite{McNeile:2011ng}. 
They also agree well with the experimental values 
of 0.4764(3) GeV and 0.6674(12) GeV~\cite{Tanabashi:2018oca}, allowing for 
the $\sim$ 3--5 MeV effect from missing QED and $\eta_b$ and $\eta_c$ 
annihilation processes 
in the lattice QCD results. 

\begin{figure}[ht]
  \includegraphics[width=0.5\textwidth]{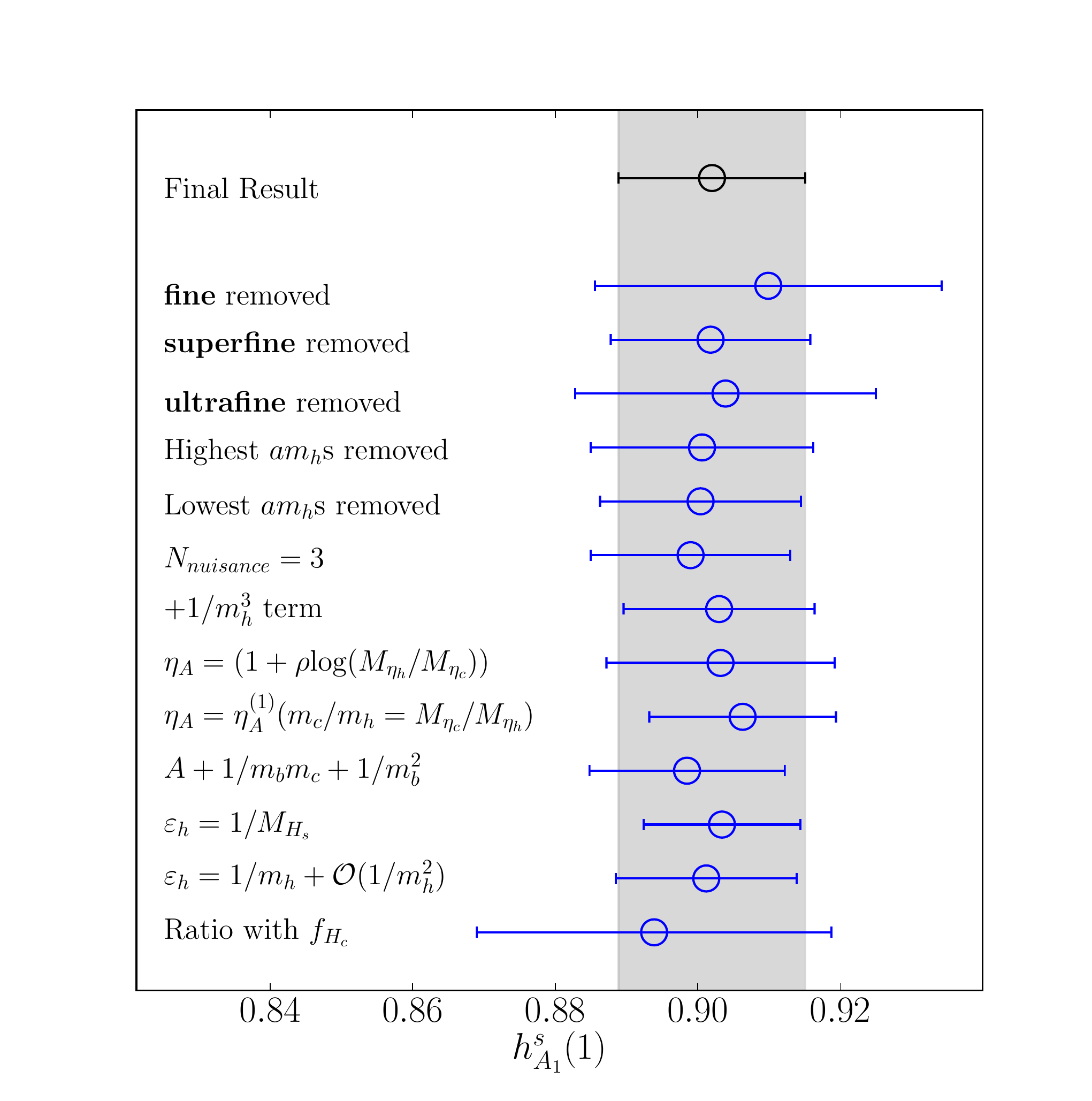}
  \caption{Results of testing the fit to $h_{A_1}^s(1)$ results. 
The top black point gives our baseline fit result in the continuum and at physical $b$ quark mass.
The top three blue points show the corresponding value 
if results from the fine, superfine or ultrafine ensembles are dropped from the fit.
    The fourth and fifth blue points show the result if instead results at the highest/lowest $am_{h0}^{\text{val}}$ value on each ensemble are removed.
    `$N_{\text{nuisance}}=3$' shows the result of truncating each sum in $\mathcal{N}_{\text{disc}}$~\eqref{eq:fitfun_hA1} at 3 rather than 2.
    `$+1/m_b^3$' results from adding an extra term to~\eqref{eq:fitform} of the form $p/M_{\eta_h}^3$ where $p$ is a fit parameter with the same 
prior as $l_{V,A,P}^s$. In this case the Bayes factor falls by a factor 
of 7, suggesting that the results do not contain a cubic dependence on the heavy mass.
    The next two points show the results of including specific 
implementations of $\eta_A$ described in 
Section~\ref{sec:extrapolation} (rather than the value 1). In the upper variant parameter $\rho$ is given prior $0\pm 1$.
The lower variant shows the result of using the 1-loop expression for $\eta_A$ (Eq.~\eqref{eq:eta-pert}), with $m_c/m_h$ replaced with $M_{\eta_c}/M_{\eta_h}$.
    `$A+1/m_bm_c+1/m_b^2$' is the result of replacing $1+l_V/m_c^2$ in the fit 
with a fit parameter $A$ with prior distribution $1\pm 1$. 
The fact that this does not affect the fit shows that mistuning of the charm quark mass is a negligible effect here.
    The points with labels beginning `$\varepsilon_h =$' show the result of replacing the heavy mass proxy $M_{\eta_h}/2$ with $M_{H_s}$ and 
the MRS quark mass (Eq.~(\ref{eq:hqet_mass})) respectively.
    The bottom point labelled `Ratio with $f_{H_c}$' is the result of an alternative 
extrapolation described in Appendix~\ref{sec:ratio_extrap}.  \label{fig:fittests}}
\end{figure}

We also performed a number of tests of our continuum/heavy-quark mass 
dependence fit to our results for $h^s_{A_1}(1)$. 
These are tabulated graphically in Figure~\ref{fig:fittests}.

One of the tests, denoted `Ratio with $f_{H_c}$' in Figure~\ref{fig:fittests}, is 
described in more detail in Appendix~\ref{sec:ratio_extrap}. 
It involves fitting the ratio of $h^s_{A_1}(1)$ to the $H_c$ decay constant, 
as a function of heavy quark mass and, after determining the continuum result 
at $m_h=m_b$, multiplying by the value for the $B_c$ decay constant determined 
from lattice QCD to obtain $h^s_{A_1}(1)$. 
The reason for doing this is because this ratio has 
smaller discretisation effects than $h^s_{A_1}(1)$ alone, as is clear from 
Figure~\ref{fig:fHcrat} in Appendix~\ref{sec:ratio_extrap}. 
It has stronger dependence on $m_h$, however, coming from the $H_c$ decay constant, 
along with sizeable uncertainties introduced from the uncertainty in the lattice spacing. 
Another disadvantage is that the physical result for $H_c$ decay constant must also be 
obtained. We find that this method gives results in agreement 
with our standard fit but with 
significantly larger uncertainties. It provides a good test, however, 
because it has very 
different $m_h$ dependence. 

\subsection{Implications for $B\to D^*$}

\begin{figure}[ht]
  \includegraphics[width=0.5\textwidth]{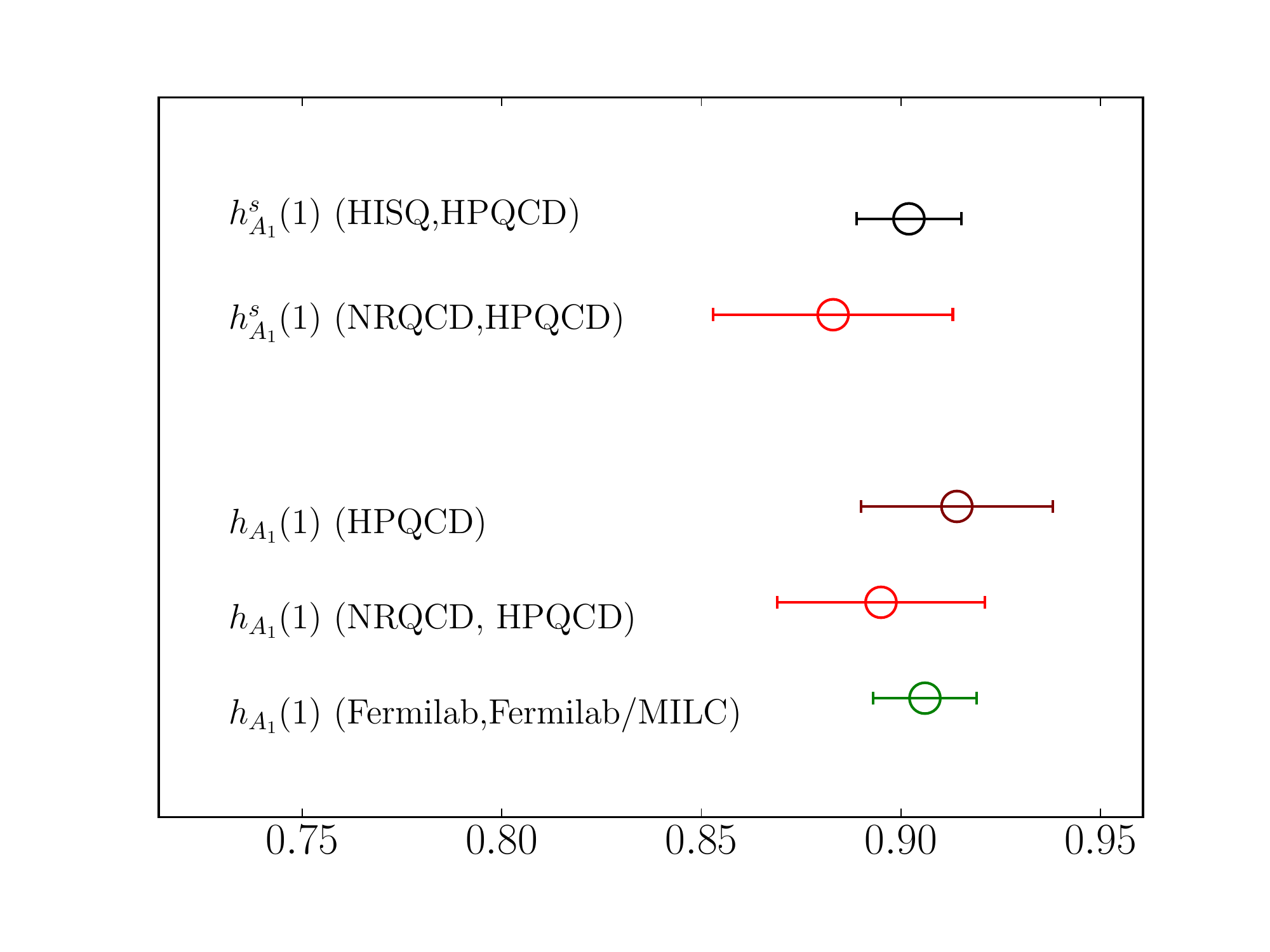}
  \caption{ Comparison of lattice QCD results for 
$h^s_{A_1}(1)$ and $h_{A_1}(1)$. 
Our results for $h_{A_1}^{(s)}(1)$ are marked `(HISQ, HPQCD)' 
and for $h_{A_1}(1)$ are marked `(HPQCD)'.  
Those marked `(NRQCD,HPQCD)' are from~\cite{Harrison:2017fmw} and the value 
marked `(Fermilab, Fermilab/MILC)' is from~\cite{Bailey:2014tva}. \label{fig:comparison}}
\end{figure}

\begin{figure}[ht]
  \includegraphics[width=0.5\textwidth]{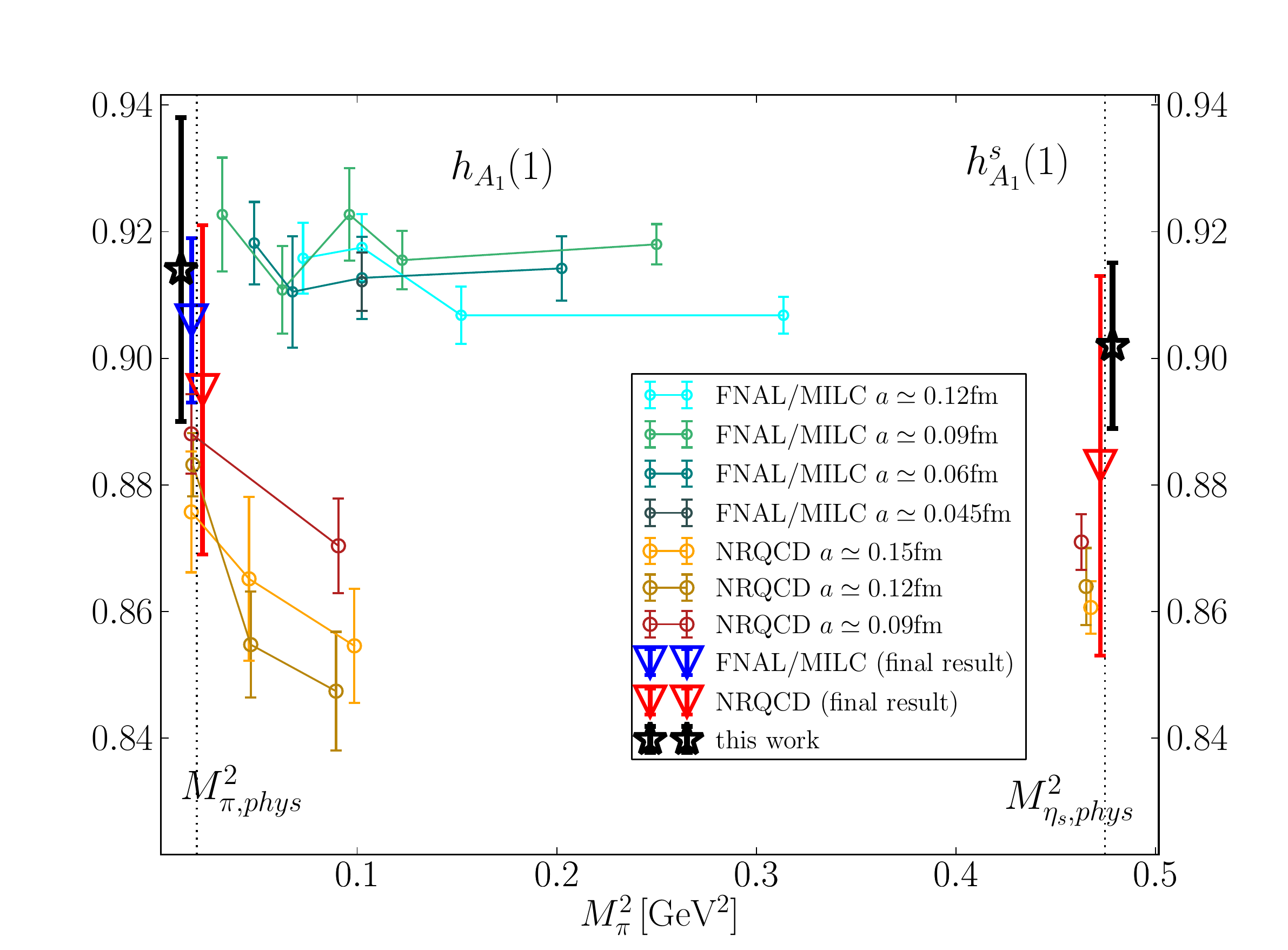}
  \caption{ More detailed comparison of lattice QCD results for 
$h_{A_1}(1)$ (left side) and $h^s_{A_1}(1)$ (right side). 
Raw results for $h_{A_1}(1)$ are
from~\cite{Harrison:2017fmw} and ~\cite{Bailey:2014tva} 
and are plotted as a function of valence 
(=sea) light quark mass, given by the square of $M_{\pi}$. 
On the right are points for $h^s_{A_1}(1)$ from~\cite{Harrison:2017fmw} 
plotted at the appropriate valence mass for the $s$ quark, but 
obtained at physical sea light quark masses.  The final result for 
$h_{A_1}(1)$ from~\cite{Bailey:2014tva}, with its full 
error bar, 
is given by the inverted blue triangle. The inverted red triangles 
give the final results for $h_{A_1}(1)$ and $h^s_{A_1}(1)$ 
from~\cite{Harrison:2017fmw}. Our results here are 
given by the black stars.  \label{fig:fermilab_nrqcd}}
\end{figure}

As discussed in Section~\ref{sec:intro}, $h^s_{A_1}(1)$ is expected to be 
close in value to the equivalent $B\to D^*$ form factor, since they only 
differ in the mass of the light spectator quark and in effects arising from the 
strong decay of the $D^*$ to $D\pi$. 
In \cite{Harrison:2017fmw} the ratio  of the two form factors was found to 
be:  $h_{A_1}(1) / h^s_{A_1}(1) = 1.013(14)_{\text{stat}}(17)_{\text{sys}}$. 
Note that systematic effects from the perturbative matching of the NRQCD-HISQ current 
largely cancel in this ratio. 

Multiplying this by our result for $h^s_{A_1}(1)$, we can determine $h_{A_1}(1)$ as
\begin{align}
\mathcal{F}^{B\to D^*}(1) = h_{A_1}(1) = 0.914(24)
\label{eq:hA1_us_nrqcd}
\end{align}
adding all the uncertainties in quadrature. 

In Figures~\ref{fig:comparison} and~\ref{fig:fermilab_nrqcd}, 
we compare current lattice results for $h_{A_1}(1)$ and $h_{A_1}^s(1)$. 
Figure~\ref{fig:comparison} compares final results for $h^s_{A_1}(1)$ from 
the HPQCD calculation using NRQCD $b$ quarks and HISQ lighter 
quarks~\cite{Harrison:2017fmw} with our full HISQ
result given here (Eq.~(\ref{eq:finalresult})). 
It also compares final results for $h_{A_1}(1)$ from using the 
Fermilab approch~\cite{Bailey:2014tva} for $b$ and $c$ quarks and 
asqtad light quarks, NRQCD $b$ 
quarks and HISQ lighter quarks~\cite{Harrison:2017fmw} 
and our result from Eq.~\ref{eq:hA1_us_nrqcd} using the strange to light ratio 
from~\cite{Harrison:2017fmw}. 
Good agreement between all results is seen, well within the uncertainties quoted. 

In Figure~\ref{fig:fermilab_nrqcd}, we show more detail of the comparison by 
plotting the lattice results from the previous Fermilab/MILC~\cite{Bailey:2014tva} 
and NRQCD $b$~\cite{Harrison:2017fmw} calculations as a function of the 
valence spectator light 
quark mass (given by the square of the pion mass). Note that, for the results for 
$h_{A_1}(1)$ to the left of the plot, the valence light and sea masses are the same. 
For the $h^s_{A_1}(1)$ points from~\cite{Harrison:2017fmw} to the right of the plot, 
the sea light (along with $s$ and $c$) quark masses take their physical values. 
Although agreement for $h_{A_1}(1)$ is seen at physical light quark mass in the 
continuum limit from all approaches, the NRQCD-HISQ results show systematic light 
quark mass dependence away from this point that is not visible in the Fermilab/MILC 
results. The two sets of results move apart as the spectator quark mass increases, 
and it is therefore not clear how well they would agree for 
spectator $s$ quarks. 

Our results, shown in Figure~\ref{fig:fermilab_nrqcd} with black stars, agree with 
the NRQCD-HISQ results for $h^s_{A_1}(1)$. The smaller uncertainties from using a 
fully nonperturbative current normalisation here show that the perturbative 
matching uncertainty allowed for 
in~\cite{Harrison:2017fmw} was conservative. Using the $s/l$ ratio from this 
calculation, where the perturbative matching uncertainty cancels, allows us to 
obtain an $h_{A_1}(1)$ result that agrees well with both earlier values. 
Our uncertainty on $h_{A_1}(1)$ is similar to that from~\cite{Harrison:2017fmw} once 
we have combined the uncertainty from the ratio with that from our value for 
$h^s_{A_1}(1)$. However we have removed the perturbative matching 
uncertainty that dominates the NRQCD-HISQ error. 

\section{Conclusions}
\label{sec:conclusions}

We have calculated the form factor at zero recoil, $\mathcal{F}^{B_s\to D^*_s}(1)$ or 
$h^s_{A_1}(1)$, using the relativistic HISQ formalism in full lattice QCD. 
This allows us to normalise the $b \rightarrow c$ current fully nonperturbatively for 
the first time and to determine how the form factor depends on the 
heavy quark mass (at physical charm quark mass). Our results show that dependence 
on the heavy quark mass is very mild (see Figure~\ref{fig:hA1_vsmetah}).   

Our result 
\begin{align}
  \mathcal{F}^{B_s\to D_s^*}(1) = h^s_{A_1}(1) = 0.9020(96)_{\text{stat}}(90)_{\text{sys}}
  \label{eq:finalresult2}
\end{align}
agrees with an earlier lattice QCD result~\cite{Harrison:2017fmw}, 
but with half the uncertainty because of the 
nonperturbative normalisation of the current. Using the strange to light 
quark ratio from the earlier paper we are able to obtain a result 
for $\mathcal{F}^{B \rightarrow D^*}(1)$  
\begin{align}
\mathcal{F}^{B\to D^*}(1) = h_{A_1}(1) = 0.914(24)
\label{eq:hA1_us_nrqcd2}
\end{align}
which is also free of perturbative matching uncertainties. 

$h^s_{A_1}(1)$ will be a useful value to compare to experimental results 
in future to determine $V_{cb}$. It has some advantages from a lattice QCD 
perspective over $h_{A_1}$ as discussed in Section~\ref{sec:intro}. 
However, $h_{A_1}(1)$  can be combined with existing experimental results to 
obtain a value for 
the CKM element $V_{cb}$. The method of combination 
has been questioned recently when it was realised that
the HQET constraints on 
the extrapolation of the exclusive experimental data to 
the zero recoil point were having a significant effect.  
Loosening these constraints
gives a higher, but less precise, value for the 
combination $|\overline{\eta}_{\text{EW}}V_{cb}|h_{A_1}(1)$ (see, for example, 
the $V_{ub}/V_{cb}$ mini-review in~\cite{Tanabashi:2018oca}). Combining this 
experimental value with lattice QCD results for $h_{A_1}(1)$ then gives a result 
for $V_{cb}$ from the $B \rightarrow D^* \ell \nu$ exclusive decay that 
agrees with, but is less accurate than, that from inclusive $b \rightarrow c$ decays. 
We do not convert our $h_{A_1}$ result into a value for $V_{cb}$ here since it is 
clear from Figure~\ref{fig:comparison} that we will agree with 
existing results (such as that in~\cite{Harrison:2017fmw}) and, on its own, our 
new result does not have sufficient accuracy to reduce uncertainties in $V_{cb}$. 

In future lattice QCD form factor calculations for both $B_s \rightarrow D_s^*$ 
and $B \rightarrow D^*$ need to 
work away from zero recoil to improve overlap with experimental results without the 
need for extrapolation\footnote{Preliminary results using the Fermilab formalism 
for $b$ and $c$ quarks and asqtad light quarks have already 
appeared~\cite{Aviles-Casco:2019vin}, as have results using M\"{o}bius domain-wall 
quarks with a range of $m_h$ values up to 2.44$m_c$~\cite{Kaneko:2018mcr}.}. 
Our results here demonstrate the efficacy of HPQCD's 
`heavy HISQ' approach for form factors at zero recoil. Away from zero recoil we 
expect it to be even more useful because it is possible to map out the full $q^2$ range 
of the decay~\cite{Lytle:2016ixw}, 
where nonrelativistic approaches must stay close to zero 
recoil because of systematic errors that grow with 
the magnitude of the daughter meson momentum. Heavy HISQ calculations are underway 
for the form factors for $B_{(s)} \rightarrow D^*_{(s)}$, 
$B_c \rightarrow J/\psi$ decay, 
and the related $B_s \rightarrow D_s$ decay over 
the full $q^2$ range, using 
the techniques developed for $c \rightarrow s$ decays to normalise the currents 
nonperturbatively~\cite{Koponen:2013tua, Donald:2013pea}.  
Initial results~\cite{Colquhoun:2016osw, McLean:2019jll} look very promising.

\subsection*{\bf{Acknowledgements}} 

We are grateful to the MILC collaboration for the use of their configurations 
and their code. Computing was done on the Cambridge Service for Data Driven Discovery 
(CSD3) supercomputer, part of which is operated by the University of Cambridge Research 
Computing Service on behalf of the UK Science and Technology Facilities Council (STFC)
DiRAC HPC Facility. The DiRAC component of CSD3 was funded by BEIS via STFC 
capital grants and is operated by STFC operations grants.  
We are grateful to the CSD3 support staff for assistance. 
Funding for this work came from STFC. 
We would also like to thank C. Bouchard, B. Colquhoun, D. Hatton, 
J. Harrison, P. Lepage and M. Wingate for useful discussions.   

\appendix

\section{Ratio method for determining $h^s_{A_1}(1)$}
\label{sec:ratio_extrap}

\begin{figure}[ht]
  \includegraphics[width=0.5\textwidth]{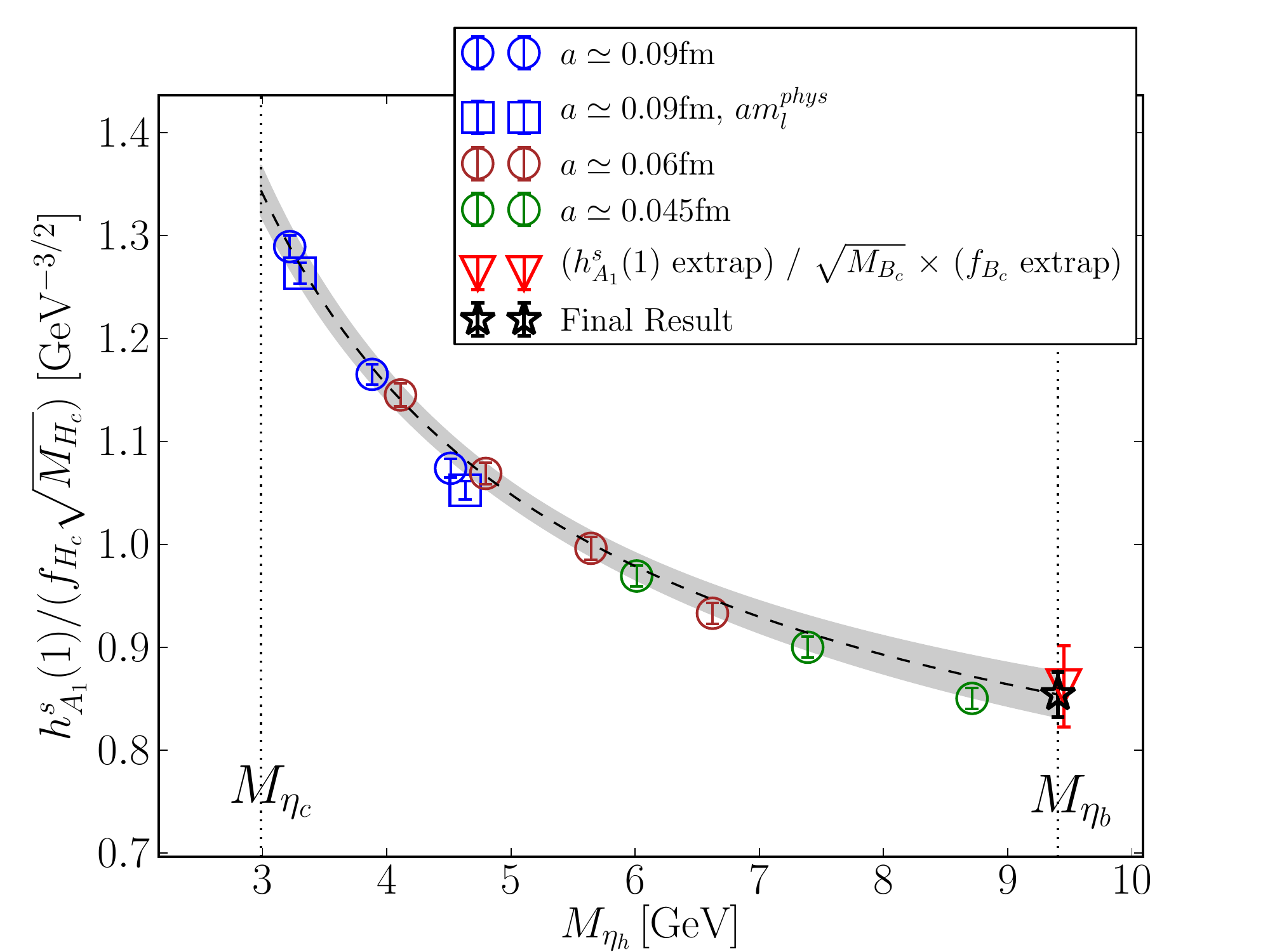}
  \caption{ The ratio $h_{A_1}^s(1)/(f_{H_c}\sqrt{M_{H_c}})$ 
plotted against $M_{\eta_h}$ (a proxy for the heavy quark mass). 
Gluon field ensembles listed in the legend follow the order of sets in 
Table~\ref{tab:ensembles}. 
The grey band shows the result of the fit described in the text, 
evaluated at $a=0$ and physical $l$, $s$ and $c$ quark 
masses to give the physical heavy quark mass dependence of the ratio. 
At $m_h=m_b$ we obtain the result given by the black star.
For comparison with our previous fit for $h^s_{A_1}(1)$ the 
inverted red triangle shows our result from Eq.~(\ref{eq:finalresult2}) 
converted to a ratio using the value for $f_{B_c}$ from Figure~\ref{fig:fHc_vsmh} 
and $M_{B_c}$ from experiment~\cite{Tanabashi:2018oca}.  \label{fig:fHcrat}}
\end{figure}

\begin{figure}[ht]
  \includegraphics[width=0.5\textwidth]{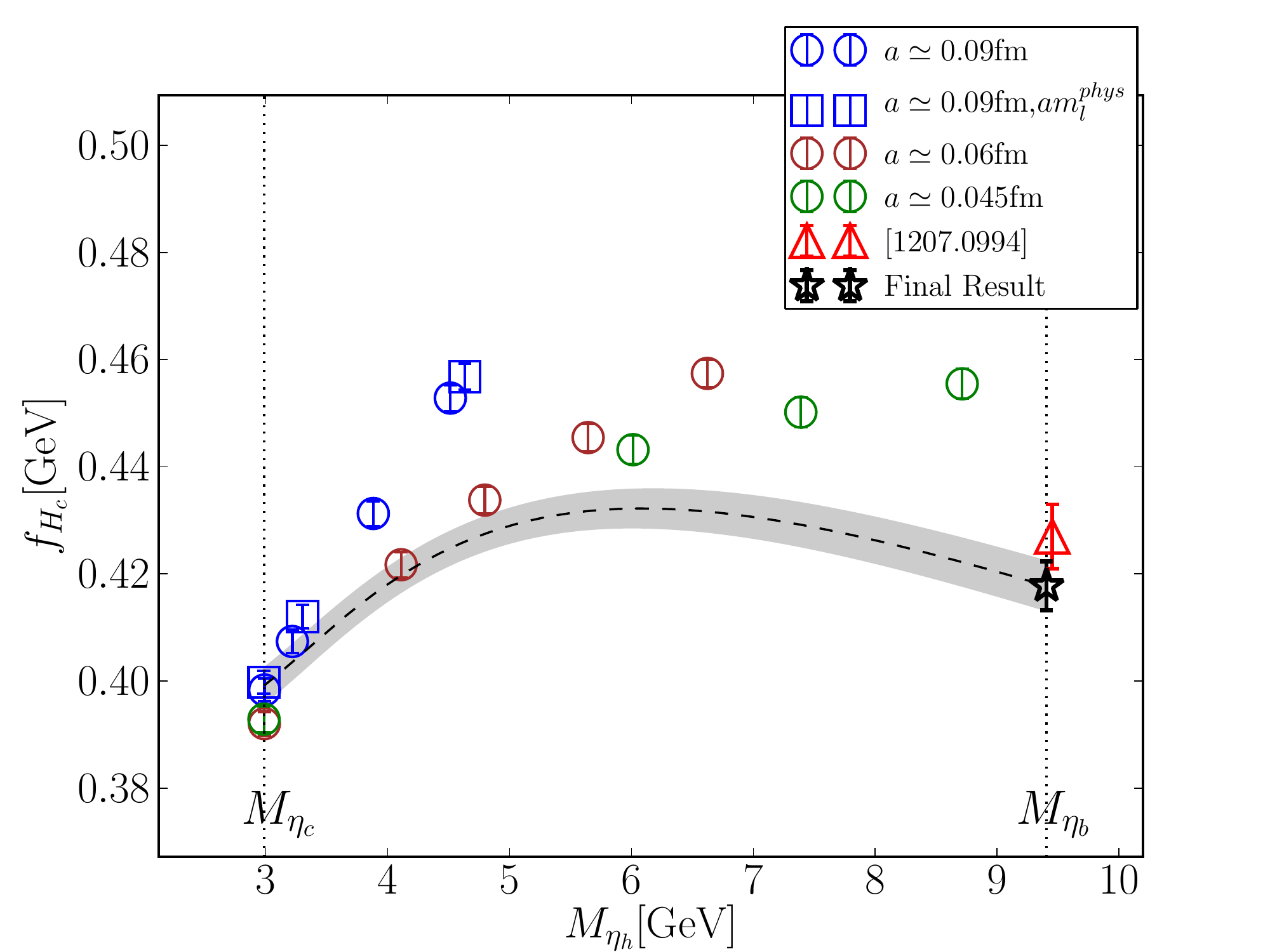}
  \caption{ The heavy-charm pseudoscalar meson decay constant, $f_{H_c}$, 
plotted against $M_{\eta_h}$ (a proxy for the heavy quark mass). 
Gluon field ensembles listed in the legend follow the order of sets in 
Table~\ref{tab:ensembles}. 
The grey band shows the result of the fit described in the text, 
evaluated at $a=0$ and physical $l$, $s$ and $c$ quark 
masses to give the physical heavy quark mass dependence of the decay constant. 
At $m_h=m_b$ we obtain the result given by the black star. 
The red triangle shows the result from a previous 
heavy HISQ determination of $f_{B_c}$ on $n_f=2+1$ gluon field 
ensembles~\cite{McNeile:2012qf}. \label{fig:fHc_vsmh}}
\end{figure}

It turns out that the significant discretisation effects visible in our results 
for $h^s_{A_1}(1)$ (Figure~\ref{fig:hA1_vsmetah}) are largely cancelled when 
we divide them by lattice QCD results for the decay constant of the 
heavy-charm pseudoscalar meson, $f_{H_c}$. 
This was also observed in~\cite{Colquhoun:2016osw} for vector form 
factors involving a $b\overline{c}$ current. 
$f_{H_c}$ is determined from the matrix 
element between the vacuum and the $H_c$ meson of the temporal axial 
vector $b\overline{c}$ current, whereas $h^s_{A_1}(1)$ is the matrix element between 
the $H_s$ and $D^*_s$ mesons of the spatial axial vector $b\overline{c}$ current. 
They behave very differently as a function of heavy quark mass but in practice have 
similar discretisation errors (compare Figures~\ref{fig:hA1_vsmetah} and~\ref{fig:fHc_vsmh}). We can make use of this in fitting the heavy quark mass dependence of their 
ratio with reduced discretisation effects. We also then need to fit the $H_c$ decay 
constant on its own in order to determine a physical value for the $B_c$ that we 
can use to determine $h^s_{A_1}(1)$ at the physical point.   

$f_{H_c}$ is found using the PCAC relation for HISQ quarks
\begin{align}
  f_{H_c} = { m_{h0} + m_{c0} \over M_{H_c}^2 } \langle 0 | P | H_c \rangle |_{\text{lat}},
  \label{eq:decayconstant}
\end{align}
where $\langle 0 | P | H_c \rangle$ is determined in the fit to the 
$H_c$ two-point correlation 
functions via~\eqref{eq:decayconstantfit}. 
We use a pseudoscalar operator, $P$, with 
spin-taste $\gamma_5 \otimes \gamma_5$ so $f_{H_c}$ is absolutely normalised. 
Results for $f_{H_c}$ for each ensemble are given in Table~\ref{tab:results} and 
plotted in Figure~\ref{fig:fHc_vsmh}. 

On each ensemble, at each heavy quark mass, we form the ratio 
$h^s_{A_1}(1)/(f_{H_c}\sqrt{M_{H_c}})$, plotted in Figure~\ref{fig:fHcrat}. 
Although discretisation effects largely cancel, the ratio varies strongly with changing 
heavy quark mass. This makes fitting this ratio as a function of heavy quark 
mass and lattice spacing very different to that 
of $h^s_{A_1}(1)$, with different systematic effects. 

We use a fit function of the same form for both 
 $h^s_{A_1}(1)/(f_{H_c}\sqrt{M_{H_c}})$ and $f_{H_c}$. 
Denoting the quantity being fit by $F$, we write (following~\cite{McNeile:2012qf}):
\begin{eqnarray}
  \label{eq:fitfun_ratio}
  F(a, m_h, m_l) &=& A \left({\alpha_s(M_{\eta_h}/2)\over\alpha_s(M_{\eta_c}/2)}\right)^p M_{\eta_h}^{n/2} \times \\ 
&&\hspace{-4.0em}\sum_{i,j,k=0}^{2,2,2} d_{ijk} \left({2\,{\text{GeV}}\over M_{\eta_h} }\right)^{i} \left({ am^{\text{val}}_{h0} \over \pi }\right)^{2j} \left({ am^{\text{val}}_{c0} \over \pi }\right)^{2k} \times \nonumber \\
  \bigg(1 &+& \mathcal{N}_{\text{mistuning}} +c_c \frac{M_{\eta_c}-M_{\eta_c}^{\text{physical}}}{M_{\eta_c}^{\text{physical}}} \bigg) . \nonumber
\end{eqnarray}
$\alpha_s(M)$ is the QCD coupling constant evaluated at scale $M$ and the 
ratio of $\alpha_s$ factors resums leading logarithms in HQET in the 
decay constant~\cite{Neubert:1993mb}. We take $\alpha_s$ in the $\overline{\text{MS}}$ 
scheme from lattice QCD~\cite{Chakraborty:2014aca}.  
The power $p$ is then -6/25 (for $n_f=4$) for 
the $f_{H_c}$ fit and +6/25 for the fit to the ratio 
$h^s_{A_1}(1)/(f_{H_c}\sqrt{M_{H_c}})$. 
The leading power of $M_{\eta_h}$, $n$, is -1 for the fit to $f_{H_c}$ based on HQET 
expectations, but 0 for the fit to the ratio because we have used 
$f_{H_c}\sqrt{M_{H_c}}$ in the denominator to remove half-integer powers of 
$\varepsilon_h$ from the fit. The remainder of the fit function allows for inverse 
powers of $m_h$ and discretisation effects. $\mathcal{N}_{\text{mistuning}}$ is the 
same as that defined earlier for our $h^s_{A_1}$ fit and is 
given in Eq.~(\ref{eq:mistuning}). The final term allows for $c$ quark mistuning 
with prior on $c_c$ of $0\pm 1$. We take a prior on the overall constant $A$ 
of $0\pm 4 (\text{GeV}^{3/2})$ in the $f_{H_c}$ fit and 
$0 \pm 2 (\text{GeV}^{-3/2})$ in the ratio fit. Priors on the 
$d_{ijk}$ are taken as $0\pm 2$ except for $d_{000}$ which is defined 
to have value 1.0. 

The fit to the ratio is shown in Figure~\ref{fig:fHcrat} and the 
fit to $f_{H_c}$ in Figure \ref{fig:fHc_vsmh}. For the ratio 
fit $\chi^2/\text{dof}$ is 0.27 for 12 degrees of freedom and for 
the $f_{H_c}$ fit, 0.53 for 16. 
Our final result for $f_{H_c}$ at $m_h=m_b$ agrees with a previous HPQCD heavy HISQ 
determination on gluon field configurations including $n_f=2+1$ flavours of sea 
quarks~\cite{McNeile:2012qf} (shown as the red triangle in 
Figure~\ref{fig:fHc_vsmh}). 
Our final result for the ratio $h^s_{A_1}(1)/(f_{H_c}\sqrt{M_{H_c}})$ at 
$m_h=m_b$ can then be multiplied by our value for $f_{B_c}$ and the square root of 
the mass of the $B_c$ meson from the Particle Data Tables~\cite{Tanabashi:2018oca}, 
to give $h^s_{A_1}(1)$. This value is shown as the bottom point in 
Figure~\ref{fig:fittests}. Figure~\ref{fig:fHcrat} compares the result from 
the ratio fit given by the grey band to the value (shown by inverted 
red triangle) obtained by taking our baseline fit result for $h^s_{A_1}(1)$ 
from Eq.~(\ref{eq:finalresult2}) and calculating from it the value of the ratio 
$h^s_{A_1}(1)/(f_{H_c}\sqrt{M_{H_c}})$ using our value for $f_{B_c}$ and the 
experimental $M_{B_c}$. The agreement is good, showing the consistency of the 
two different approaches.

\bibliography{paper}

\end{document}